\documentclass{aa}  
\usepackage{graphicx}
\graphicspath{{figures/}}
\usepackage[]{hyperref}
\usepackage{xcolor}
\usepackage{float}
\usepackage{subcaption}
\usepackage{makecell}
\usepackage{multirow}

\hypersetup{
  colorlinks   = true, 
  urlcolor     = blue, 
  linkcolor    = blue, 
  citecolor   = blue 
}

\newcommand{\Swift}{\textit{Swift}}
\newcommand{\HST}{\textit{HST}}
\newcommand{\SN}{SN~2024abfo}

\usepackage{txfonts}
\begin{document}

   \title{A low mass, binary-stripped envelope for the Type IIb SN~2024abfo}

   \author{S. de Wet\inst{1} \and G. Leloudas\inst{1} \and D. Buckley\inst{2,3} \and N. Erasmus\inst{2,4} \and P.J. Groot\inst{2,3,5} \and E. Zimmerman\inst{6}
          }

   \institute{DTU Space, Technical University of Denmark, Building 327, Elektrovej, 2800 Kgs. Lyngby, Denmark 
              \and 
              South African Astronomical Observatory, PO Box 9, 7935 Observatory, South Africa
 \and
 Department of Astronomy, University of Cape Town, Private Bag X3, Rondebosch 7701, South Africa
\and
Department of Physics, Stellenbosch University, Stellenbosch, 7602, South Africa
\and
Department of Astrophysics/IMAPP, Radboud University, PO Box 9010, 6500 GL Nijmegen, The Netherlands
\and
Department of Particle Physics and Astrophysics, Weizmann Institute of Science, 76100 Rehovot,
Israel
             }

   \date{Latest draft: \today}

 
  \abstract
   {Type IIb supernovae (SNe) are a transitional subclass of stripped-envelope SNe showing hydrogen lines in their spectra that gradually weaken and give way to helium lines reminiscent of SNe Ib. The presence of hydrogen indicates that they retain a non-negligible hydrogen-rich envelope that has been stripped through stellar winds or binary interaction. }
   {The direct detection of SN progenitors is a valuable way to connect the various supernova sub-types with their progenitor stars. SN~2024abfo is the seventh SN IIb with a direct progenitor detection. Our aim is to study the progenitor candidate and the supernova itself to determine the evolutionary history of the system. }
   {We astrometrically-register our ERIS adaptive optics imaging with archival HST imaging to determine whether the supernova position is consistent with the progenitor candidate position. We perform photometry on archival DECam imaging to derive the spectral energy distribution of the progenitor candidate and investigate its temporal variability. We consider single and binary star models to explain the end point of the progenitor candidate in the Hertzsprung-Russell diagram. For the supernova we compare the light curves and spectra with other SNe IIb with progenitor detections. We derive the bolometric light curve and attempt to fit this with a semi-analytic light curve model.}
   {The position of the supernova in our adaptive optics imaging agrees with the progenitor position to within 19~mas. The progenitor SED is consistent with an A5 giant with a radius of ${\sim}125~R_\odot$, a temperature of ${\sim}8400$~K, and a luminosity of $\log(L/L_\odot){\sim}4.8$. Single star models predict an initial mass in the range 11--15~$M_\odot$, while the most probable binary model is a $12+1.2~M_\odot$ system with an initial period of 1.73 years. We also find significant evidence for variability of the progenitor candidate in the years prior to core-collapse. SN~2024abfo is the least luminous SN IIb with a direct progenitor detections. At late times the $r$-band light curve decays more slowly than the comparison SNe, which may be due to increased $\gamma$-ray trapping, although this requires further investigation. Similar to SN~2008ax, SN~2024abfo does not show a prominent double-peaked light curve. Our semi-analytic light curve modelling shows that this may be due to a very low mass of hydrogen (${\lesssim}0.0065~M_\odot$) in the outer envelope. Spectrally, SN~2024abfo is most similar to SN~2008ax at early times while at later times (${\sim}80$~days) it appears to show persistent H$\alpha$ absorption compared to the comparison sample. }
   {We prefer a binary system to explain \SN{} and its progenitor, although we are unable to rule out single-star models. We recommend late-time observations to search for a binary companion and signatures of CSM-interaction. The absence of these features would support the hypothesis that \SN{} resulted from a system which underwent a period of binary mass transfer well before ($\gtrsim1000$~yr) the explosion, resulting in a low-mass ($\lesssim0.01~M_\odot$) hydrogen-rich envelope. }


   \maketitle
%

\section{Introduction} \label{sec:intro}
Core-collapse supernovae (SNe) are violent explosions marking the end of the lives of massive stars with initial masses ${>}8~M_\odot$ \citep{Smartt2009}. The evolution of these stars in the final stages prior to explosion remains one of the leading open questions in stellar astrophysics. 
Type IIb SNe are a transitional spectroscopic class which initially show hydrogen features that gradually weaken and give way to helium lines characteristic of SNe Ib \citep{Filippenko1997}. This indicates that they have retained a small but non-neglible hydrogen-rich envelope, making them valuable objects for studying how stripped-envelope SNe (of types IIb/Ib/Ic) lose their hydrogen envelopes. 
 
The direct detection of SN progenitor stars in pre-explosion imaging offers a powerful means of connecting the various SN spectroscopic classes with their progenitor stars \citep[for reviews see][]{Smartt2009,Smartt2015,Vandyk2017}. A small but growing number of such progenitors have been identified in high-resolution archival imaging, primarily from the \textit{Hubble Space Telescope} (\textit{HST}). The majority of these (${\sim}20$) are red supergiants associated with Type IIP SNe \citep[e.g. SN~2024ggi;][]{Xiang2024}, with inferred masses in the range 8--18~$M_\odot$ \citep{Smartt2015}. A smaller number of Type IIb SNe have progenitor detections, with six confirmed cases: 1993J \citep{Aldering1994,Maund2004}, 2008ax \citep{Crockett2008,Folatelli2015}, 2011dh \citep{Maund2011dh,Vandyk2011dh}, 2013df \citep{Vandyk2014}, 2016gkg \citep{Kilpatrick2017,Kilpatrick2022}, and 2017gkk \citep{Niu2024}. A wide variety of progenitor properties have been deduced from these studies, with spectral types ranging from K to B supergiants (for SNe 1993J and 2008ax, respectively), radii from 40--550~$R_\odot$, and estimated main sequence masses between 10 and 20~$M_\odot$.  

The photometric and spectral properties of Type IIb SNe show a similar diversity. Some have double-peaked light curves (e.g SNe~1993J, 2013df, 2016gkg) while others do not (e.g. SNe~2008ax, 2011dh). The first peak has been interpreted as cooling emission from the hydrogen-rich envelope following shock breakout \citep{Blinnikov1998,Nakar2010,Sapir2017} while the second peak is powered by radioactive heating from \textsuperscript{56}Ni. The presence of the H$\alpha$ line in spectra is the primary diagnostic feature used to differentiate IIb and Ib SNe. Estimates of the amount of H in the envelope required to produce a clear H$\alpha$ line vary from as little as 0.001~$M_\odot$ \citep{Dessart2011} to a few 0.1~$M_\odot$ \citep{Dessart2018}. Modelling has shown that objects on the higher mass end of this distribution have a prominent shock-cooling peak \citep[e.g. 0.2~$M_\odot$ for SN~1993J;][]{Woosley1994} while for those with lower masses the peak may be weaker or absent entirely \citep{Dessart2011,Bersten2012}. The envelope mass is therefore the key factor in determining whether light curves are double-peaked or not. This hypothesis is supported by \citet{Chevalier2010}'s division of IIb SNe into 'compact' (cIIb, $R_*{\sim}10^{11}$~cm) and 'extended' (eIIb, $R_*{\sim}10^{13}$~cm) sub-categories with a dividing envelope mass of ${\sim}0.1~M_\odot$ separating the two classes. They suggested that cIIb SNe have weak or absent shock-cooling emission (e.g. SN~2008ax), weak H$\alpha$ emission during the nebular phase, rapidly evolving radio emission, non-thermal X-ray emission, and arise from Wolf-Rayet progenitors. Based on radio and X-ray observations of SN~2011dh \citet{Soderberg2012} preferred the cIIb scenario, although this contrasted with the yellow supergiant identified in pre-explosion imaging \citep{Vandyk2011dh} and hydrodynamical modelling which found a radius of ${\approx}200~R_\odot$ \citep{Bersten2012}, casting doubt on the compact versus extended framework. 

One aspect of Type IIb SNe where there appears to be more agreement is the mechanism of stripping of the H-rich envelope. Although single-star models can reproduce Type IIb SNe \citep[e.g. ][]{Groh2013}, binary models are favoured due to the direct detection of the binary companions for SNe 1993J \citep{Maund2004,Fox2014}, 2011dh \citep{Folatelli2014,Maund2019} and 2001ig \citep{Ryder2018}. Also, the fact that more than 70\% of massive stars are in binary systems which undergo mass transfer \citep{Sana2012} makes it statistically more probable that IIb progenitors are formed in binaries. In a study of SN 2013df \citet{Maeda2015} put forward the intriguing hypothesis that the timing of a strong binary interaction could explain the differences in mass-loss rate and extent of the progenitor star's envelope, with extended-envelope SNe (e.g. 1993J and 2013df) having had a more recent binary interaction compared to compact-envelope SNe (such as 2008ax and 2011dh). Direct progenitor detections and subsequent searches for the remaining binary companion offer a unique way to test the compact vs extended progenitor question and the timing of a putative binary interaction. 


Here we report on the Type IIb \SN{}, the seventh Type IIb SN with a progenitor detection. In Section \ref{sec:discovery} and \ref{sec:observations} we present the discovery of \SN{} and our observations. In Section \ref{sec:progenitor} we study the progenitor candidate using archival imaging with the aim of deriving its basic properties and evolutionary history. In Section \ref{sec:supernova} we present our photometric and spectroscopic dataset on \SN{} and compare its properties to those of other Type IIb SNe with progenitor detections. In Section \ref{sec:discussion} we discuss the implications of our findings. All magnitudes are in the AB system unless stated otherwise. 




\section{Discovery and host galaxy}\label{sec:discovery}
SN~2024abfo was discovered by the Asteroid Terrestrial-impact Last Alert System (ATLAS) on 2024 November 15 at 22:54:00~UT \citep{2024TNSAN.341}. Following the ATLAS discovery, spectroscopic observations of the new transient were automatically triggered with the Mookodi low-resolution spectrograph and imager. We classified the object as a Type II supernova due to the presence of a broad P-Cygni H$\alpha$ feature in the spectrum \citep{2024TNSAN.342}. In addition to the optical detections, SN 2024abfo was detected in X-rays by the X-Ray Telescope (XRT) aboard \Swift{} \citep{2024TNSAN.343} and at radio frequencies by the Australia Telescope Compact Array \citep[ATCA;][]{ATCA_atel}. The coordinates\footnote{\href{https://www.wis-tns.org/object/2024abfo}{https://www.wis-tns.org/object/2024abfo}} of \SN{} are $\alpha=03^\mathrm{h}57^\mathrm{m}25.62^\mathrm{s}$, $\delta=-46^\circ11^{\prime}07.67^{\prime\prime}$ (J2000). ATLAS obtained four observations of the field containing the supernova on the night of its discovery. The first $3\sigma$ detection occurred during the third observation of the night, thereby constraining the explosion epoch to within 24.6 minutes of the previous non-detection which occurred at the Modified Julian Date (MJD) of 60628.28 \citep{2024TNSAN.341}. We reference all times with respect to this epoch henceforth. 

The host galaxy of SN~2024abfo is the barred spiral galaxy NGC 1493. The NASA/IPAC Extragalactic Database\footnote{\href{https://ned.ipac.caltech.edu}{https://ned.ipac.caltech.edu}} (NED) gives an SB(r)cd morphological classification for NGC 1493, and lists a total of six redshift-independent distances for NGC 1493 spanning 9.25--12.70~Mpc, all-derived using the Tully-Fisher relation. Taking the mean and standard deviation of these six estimates we derive a distance to NGC 1493 of $10.85\pm1.29$~Mpc and an associated distance modulus of $\mu=30.17\pm0.26$. We adopt the fiducial host galaxy redshift of $z=0.003512$ as reported on NED henceforth. The Milky Way dust reddening toward this line of sight is small, with $E(B-V)=0.0089$~mag according to the dust maps of \citet{SF2011}. We follow \citet{Reguitti2025} in assuming negligible host galaxy extinction due to the lack of \ion{Na}{i} D absorption lines at the host galaxy redshift. \citet{Rossa2006} used \HST{} spectroscopic observations to study nuclear star clusters in a number of spiral galaxies, one of which was NGC~1493. Through spectral population fitting they found negligible extinction and solar metallicity ($Z=0.02$) for the nuclear cluster in NGC 1493. The metallicity at the site of \SN{} is therefore likely to be solar or sub-solar due to a metallacity gradient which is often observed in spiral galaxies.     

\section{Observations}\label{sec:observations}
\subsection{Adaptive optics imaging}
We obtained adaptive optics (AO) imaging of \SN{} with the Enhanced Resolution Imaging Spectrometer \citep[ERIS;][]{Davies2023} mounted at the Cassegrain focus of UT4 at the Very Large Telescope, through director's discretionary time proposal 114.28HG.001 (PI: de Wet). Our observations were obtained on 2025 February 2 and consisted of a total of 20 minutes of exposure in the $J$ band with the NIX imager, employing the lower resolution (27 versus 13 mas) pixel scale to ensure a larger ($55^{\prime\prime}$ versus $26^{\prime\prime}$) field-of-view. The larger field-of-view was chosen so that there would be sufficient objects for astrometrically registering the ERIS image with the archival HST image. The 20 minutes of exposure were divided into 10 dithers in order to alleviate the effect of two large areas of bad pixels on the NIX detector. Each dither consisted of 60 detector integrations using an integration time of 2~s. The data was read out using the slow readout mode and the full-frame $2048\times2048$ pixel array. Due to the brightness of the supernova at the time of the observations ($z{\sim}15$ mag) we used the Laser Guide Star (LGS) AO mode with the supernova itself as the tip-tilt star to correct for atmospheric distortions. We reduced the ERIS/NIX data with the \verb|eris_nix_img| workflow within the \verb|EsoReflex| environment. The workflow corrects observed frames for the dark and sky background, flatfields the data, performs astrometric and photometric calibration, and stacks the individual dithers. The output of the pipeline is a final stacked image with a 27~mas pixel scale and a catalogue file containing all detected sources above a user-specified threshold. 

\subsection{Photometry}\label{subsec:phot}
We obtained photometry with the Mookodi low-resolution spectrograph and imager \citep{Erasmus2024} mounted on the Lesedi 1~m telescope \citep{Worters2016} situated at the South African Astronomical Observatory site in Sutherland, South Africa. Observations consisted of 60~s exposures in each of the Sloan Digital Sky Survey (SDSS) $griz$ filters. Gaps in our sampling of the light curve were due to poor weather at Sutherland or scheduling constraints. We used an adapted version of the MeerLICHT and BlackGEM pipelines \citep[Vreeswijk et al., in preparation;][]{Groot2024} to compute the supernova photometry. The pipeline computes photometry using the optimal extraction method after determining the image point-spread-function (PSF) from a number of bright, unsaturated stars in the field-of-view (FOV). The image zero point is calculated from stars in the FOV which are listed in a dedicated Mookodi calibration catalogue. This calibration catalogue was created using Gaia DR3 sources and takes into account the Mookodi filter sensitivity curves. 

The \Swift{} Ultra-violet Optical Telescope \citep[UVOT;][]{Roming2005} observed SN~2024abfo as a result of target-of-opportunity (ToO) requests from a number of users (including Sand, Campana, Zimmerman, de Wet, Farah, and Pazhayath Ravi). A total of 28 epochs---designated by separate Obs IDs---were obtained spanning the ultra-violet ($uvm2$, $uvw1$, $uvw2$, $U$) and optical ($B$, $V$) bands. We downloaded all publicly available UVOT photometry from the Swift Archive Download portal hosted on the UK Swift Science Data Centre website\footnote{\href{https://www.swift.ac.uk/swift_portal/}{https://www.swift.ac.uk/swift\_portal/}}. We used the HEASoft Swift FTOOLS software package, version 6.34\footnote{\href{https://heasarc.gsfc.nasa.gov/docs/software/heasoft/}{https://heasarc.gsfc.nasa.gov/docs/software/heasoft/}} to process the level 2 sky images. For certain epochs, more than one exposure was obtained in a particular filter. In these cases we used the \verb|uvotimsum| tool to coadd exposures on a per-epoch basis. Thereafter we used the \verb|uvotsource| tool to perform aperture photometry on the coadded images, where the source and background apertures were set to $5^{\prime\prime}$ and $10^{\prime\prime}$ in radius, respectively. We report $3\sigma$ limiting magnitudes in cases where the signal-to-noise (SNR) ratio of the supernova was below three.

We obtained ATLAS $o$-band photometry of SN 2024abfo by querying the ATLAS Forced Photometry server\footnote{\href{https://fallingstar-data.com/forcedphot/}{https://fallingstar-data.com/forcedphot/}} \citep{Tonry2018,Smith2020}. We do not use the $c$-band data since it is sparsely sampled compared to the $o$-band. We require detections to be at the $3\sigma$ level, while for non-detections we report $3\sigma$ upper limits. We present all photometry for SN 2024abfo in Table \ref{tab:SN_phot}. 

\subsection{Spectroscopy}
Along with our photometric observations, we also obtained regular spectroscopic observations with Lesedi/Mookodi. Mookodi has a stepped-slit configuration with narrow ($2^{\prime\prime}$) and wide ($4^{\prime\prime}$) slit widths corresponding to spectral resolutions of $R\approx350$ and $R\approx175$ at $6000~\AA$, respectively \citep{Erasmus2024}. Due to Mookodi's fully-robotic operation, the wider slit is usually used for observations of transients in order to ensure that the majority of the flux is captured. Each observation consisted of a 600~s exposure of the target followed by a 4~s arc frame for wavelength calibration, all employing the wide slit. We reduced the data using a Mookodi-specific pipeline adapted from the open-source, python-based spectral reduction software toolkit known as Automated SpectroPhotometric REDuction \citep[ASPIRED;][]{Lam2023}. The pipeline performs cosmic-ray removal, trace fitting and optimal spectral extraction, wavelength calibration, and flux calibration using observations of a spectrophotometric standard star. 

We additionally obtained six epochs of spectra with the Southern African Large Telescope \citep[SALT;][]{Buckley2006} equipped with the long-slit Robert Stobie Spectrograph (RSS) through observing program 2024-2-LSP-001 (PI Buckley). Observations consisted of 1200~s of exposure using the PG0700 grating with a grating angle of 4.60 degrees resulting in a wavelength range of 3590--7480~\AA\ with a spectral resolution of $R=735$ at the central wavelength. Standard SALT data products provided to users include a cosmic-ray cleaned, bad-pixel corrected, and wavelength-calibrated 2D spectrum. We performed optimal spectral extraction of the supernova trace using the ASPIRED toolkit, and flux-calibrated the spectra via observations of a spectro-photometric standard star.

\section{The progenitor candidate}\label{sec:progenitor}
\subsection{Image registration}
\begin{figure*}
\centering
\includegraphics[width=0.8\textwidth]{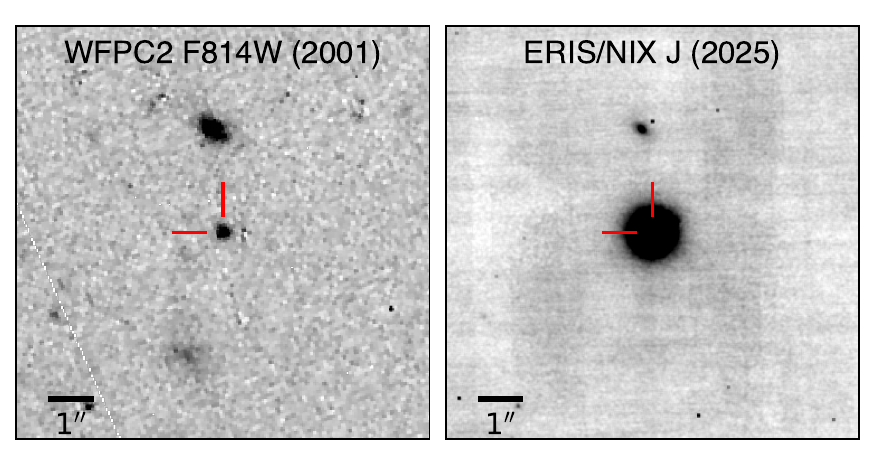}
\caption{Archival \textit{Hubble Space Telescope}/WFPC2 F814W image from 2001 (left) and our ERIS/NIX $J$-band image (right) of \SN{} obtained in February 2025. The images have been astrometrically-registered to a common reference frame using 10 sources visible in both images. Both images are $10^{\prime\prime}\times10^{\prime\prime}$ in dimension, with north up and east to the left. The red crosshairs indicate the position of \SN{} as identified in our ERIS image. The ERIS pixel scale is 27~mas per pixel, while the drizzled WFPC2 image has a pixel scale of 45.5~mas per pixel. The supernova position agrees with the progenitor position within the 19~mas uncertainty of the coordinate transformation between the two frames. }
\label{fig:ERIS}
\end{figure*}

The \textit{Hubble Space Telescope (HST)} fortuitously observed the location of \SN{} in 2001 as part of \HST{} proposal 8599 targeting nuclear star clusters of late-type spiral galaxies \citep{Boeker2002}. A single 640~s observation of the supernova field was taken with the Wide Field Planetary Camera 2 (WFPC2) in the F814W filter, and consisted of three individual exposures of 40, 300 and 300~s, respectively. The Hubble Legacy Archive\footnote{\href{https://hla.stsci.edu}{https://hla.stsci.edu}} drizzled image had a pixel scale of $0.1^{\prime\prime}$ per pixel and an incorrect astrometric solution. We therefore retrieved the raw and processed data from the Barbara A. Mikulski Archive for Space Telescopes (MAST) via the Hubble search form\footnote{\href{https://mast.stsci.edu/search/ui/\#/hst}{https://mast.stsci.edu/search/ui/\#/hst}}. The final combined and drizzled image from the MAST archive had a finer pixel sampling of $0.045^{\prime\prime}$ per pixel and a better astrometric solution due to a more recent processing. We ran \verb|SExtractor| on the data to extract the positions of sources and identified 10 objects common to both the WFPC2 and ERIS images in order to calculate a linear coordinate transformation between the two frames. The RMS uncertainty of the transformation was 0.43 WFPC2 pixels, or 19~mas. The position of \SN{} in the ERIS image was determined to an accuracy of 0.5~mas, while the position of the progenitor candidate in the WFPC2 image was measured to 1.5~mas accuracy. Transforming the ERIS position to the WFPC2 pixel frame resulted in a position that was 0.10 WFPC2 pixels (4.6~mas) away from the progenitor candidate position  (see Figure \ref{fig:ERIS}). Considering the much larger uncertainty of 19~mas due to the coordinate transformation, we consider the position of the progenitor candidate to be fully consistent with the position of \SN{} and it is therefore a strong candidate as the actual progenitor star. We caution, however, that the progenitor candidate we have identified will remain only a candidate until its disappearance is confirmed via late-time high-resolution images once the supernova has faded. An important case study is the Type IIb SN 2008ax for which a single progenitor candidate was identified in pre-explosion \HST{}/WFPC2 data by \citet{Crockett2008} and which was later shown to be the combined light from four unresolved sources \citep{Folatelli2015}. There is also the case of the Type Ic SN 2017ein for which a progenitor candidate was identified in pre-explosion imaging and later shown to not have disappeared \citep{Vandyk2018,Kilpatrick2018,Zhao2025}.  

\subsection{Photometry}\label{subsec:photometry}
In addition to the archival \HST{} image, we also retrieved archival imaging of the supernova site taken with the Dark Energy Camera \citep[DECam;][]{DECAM} mounted on the Victor M. Blanco 4~m telescope. Crucially, the DECam imaging encompassed the $grizY$ bands, which allows us to perform a far more comprehensive analysis of the progenitor SED than with the \HST{} data alone. For the DECam data we queried the NOIRLab Astro Data Lab \citep{Fitzpatrick2014,Nikutta2020} image cutout service\footnote{\href{https://datalab.noirlab.edu/sia.php}{https://datalab.noirlab.edu/sia.php}} for stack images at the supernova position and downloaded images in the $grizY$ bands. The stacked images were created from individual 90~s exposures obtained over the course of the Dark Energy Survey (DES), and released to the public as part of Data Release 1 \citep[DR1;][]{DES2018}. The DR1 stack images comprised data taken between August 2013 and February 2016. We also obtained near-infrared $JHK$ archival imaging of the progenitor location taken with the VISTA InfraRed CAMera (VIRCAM) mounted on the 4~m Visible and Infrared Survey Telescope for Astronomy (VISTA). The processed VIRCAM data was retrieved from the ESO Archive Science Portal. No source was visible at the progenitor location, so we report $5\sigma$ upper limits provided in the image headers in Table \ref{tab:Progenitor}. As shown in Section \ref{subsec:singlestar}, the VIRCAM upper limits are not deep enough to meaningfully constrain progenitor star SED models, but we choose to include them here for completeness.

For the \HST{} data we perform point-spread function (PSF) photometry using the DOLPHOT \citep{Dolphin2000,Dolphin2016} software package with the WFPC2 module. The individual exposure images obtained from the MAST archive had been pre-processed using the standard WFPC2 pipeline. Before running DOLPHOT we applied the bad pixel mask to each image and used the task \verb|crrej| to identify and mask cosmic rays. We ran DOLPHOT using the precomputed WFPC2 PSFs and set FitSky=3, RAper=8, and InterpPSFlib=1, as in \citet{VanDyk2019}. We also set WFPC2useCTE=1 to apply charge transfer efficiency (CTE) corrections. No differences in the photometry were found while experimenting with different aperture sizes from 3 to 8 pixels and turning aperture corrections on or off. We measure an F814W (Vega) magnitude of $21.86\pm0.03$ for the progenitor candidate, which is equivalent to $22.30\pm0.03$ in the AB system. 

\begin{figure*}
\centering
\includegraphics[width=0.9\textwidth]{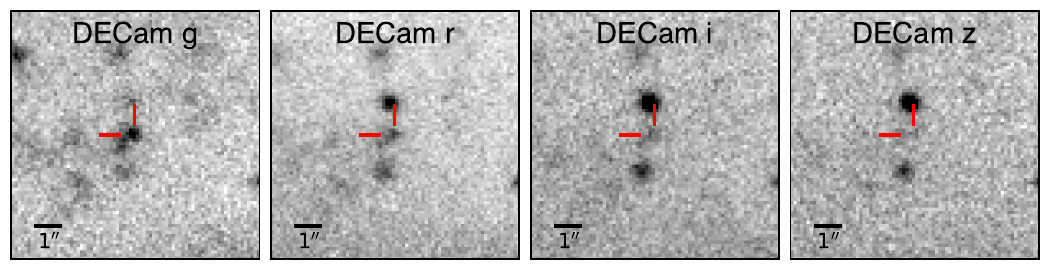}
\caption{Archival DECam stack images of the site of \SN{}. Each cutout is $20^{\prime\prime}\times 20^{\prime\prime}$ in dimension. The red crosshairs indicate the position of the progenitor candidate, with North up and East to the left.}
\label{fig:cutouts}
\end{figure*}

For the DECam data we use two approaches to compute the photometry: aperture photometry and point-spread-function (PSF) photometry. For aperture photometry we use the \verb|photutils| \citep{photutils} Python package to extract background-corrected aperture magnitudes of all point sources in the DECam stack images, with the aperture radius set to 3-pixels. We perform photometric calibration using the $grizY$ magnitudes\footnote{We extract the \texttt{mag\_auto} magnitudes from the \texttt{des\_dr1.main} table.} of matching sources listed in the DES DR1 catalogue. The photometric calibration zero points determined in this way are accurate to within 0.02 magnitudes for each band. We use a small aperture with a 3-pixel radius in order to exclude the flux contribution from a nearby source to the south east of the progenitor candidate (Figures \ref{fig:ERIS} and \ref{fig:cutouts}). For the position of the progenitor aperture we use the most accurate position we have available---the supernova position from our ERIS observations. We use a large, 30-pixel radius background region centred on the supernova position to determine the background in the vicinity of the progenitor. To exclude any contribution from the progenitor candidate (and other sources), we use sigma-clipping when calculating the background flux level. To perform PSF photometry we use the \verb|photutils.psf| subpackage to first build an effective PSF from a large number of isolated point sources in the field of view. We then derive a PSF-fitting image zero point by extracting PSF magnitudes of sources in our image and cross-matching them to the DES DR1 catalogue. We perform forced PSF photometry within a $9\times9$-pixel grid at the ERIS position while taking into account the background brightness in the vicinity of the progenitor. Our aperture and PSF photometry results show excellent agreement in the $g$ and $r$ bands, which is understandable since the SNR is highest in these bands. In the $i$ and $z$ bands the PSF magnitudes are roughly 0.1 magnitudes fainter than the aperture magnitudes, though they still agree within the $1\sigma$ errors. We choose to report the aperture photometry magnitudes in Table \ref{tab:Progenitor}. For the $Y$-band image we report a $5\sigma$ upper limit. All photometry for the progenitor candidate is presented in Table \ref{tab:Progenitor}.  

\begin{table*}
\caption{\SN{} progenitor candidate photometry}
\label{tab:Progenitor}
\centering
\begin{tabular}{l c c c c r}
\hline\hline
Telescope/Instrument & Band & $\lambda_\mathrm{eff}$ (nm) & Observation Date\tablefootmark{a} & Exposure time (s) & AB Magnitude \\
\hline
\HST{}/WFPC2       & F814W   & 793   & 2001-05-02      & 640 & $22.30 \pm 0.03$ \\
Blanco 4m/DECam    & $g$     & 477   & Stack           & 540 & $22.83 \pm 0.07$ \\
Blanco 4m/DECam    & $r$     & 637   & Stack           & 540 & $22.76 \pm 0.08$ \\
Blanco 4m/DECam    & $i$     & 777   & Stack           & 540 & $23.00 \pm 0.11$ \\
Blanco 4m/DECam    & $z$     & 915   & Stack           & 630 & $23.20 \pm 0.17$ \\
Blanco 4m/DECam    & $Y$     & 989   & Stack           & 360 & $>22.34$ \\
VISTA/VIRCAM       & $J$     & 1248  & 2011-12-09      & 360 & $>20.56$ \\
VISTA/VIRCAM       & $H$     & 1635  & 2011-12-09      & 360 & $>20.05$ \\
VISTA/VIRCAM       & $K_s$   & 2144  & 2011-12-09      & 360 & $>19.97$ \\
\hline
\end{tabular}
\tablefoot{
\tablefoottext{a}{The stacked DECam images were created from exposures taken between August 2013 and February 2016.}
}
\end{table*}

\subsection{Variability}\label{subsec:variability}
The DECam $i$-band and WFPC2/F814W magnitudes in Table \ref{tab:Progenitor} differ by almost one magnitude, despite the two filters having very similar effective wavelengths. As a check on our \HST{} photometry we performed aperture photometry on a number of bright stars in the WFPC2 image and derived an image zero point calibrated to the DES DR1 $i$-band catalogue. The photometry we measure for the progenitor candidate is fully in agreement with the DOLPHOT measurement in Table \ref{tab:Progenitor}, which indicates that the progenitor candidate was indeed 0.7 magnitudes brighter in the $i$-band in May 2001 compared to 2013--2016. 

\begin{figure}
\centering
\includegraphics[width=\columnwidth]{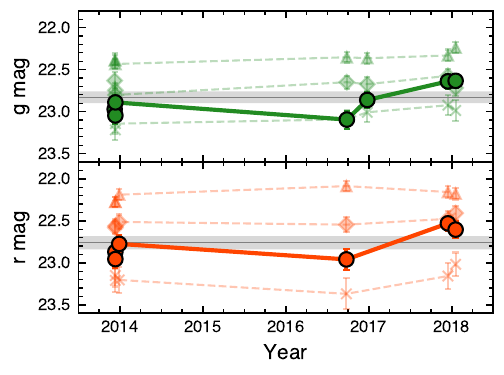}
\caption{Photometry of the progenitor candidate (solid lines) versus three nearby point sources (dashed lines) computed from individual 90~s DECam images spanning 2013 to 2018. The grey horizontal lines denote the brightness of the progenitor candidate as computed from the stack images. The grey region represents the $1\sigma$ uncertainty on the stack image photometry for the progenitor candidate.  }
\label{fig:variability}
\end{figure}

We additionally use the individual 90~s DECam images to search for variability in the $g$ and $r$ bands. We exclude the $i$ and $z$ bands due to the much lower SNR in these bands. We obtained the data from the NOIRLAB Astro Data Lab image cutout service and computed photometry using the same procedure as for the stack images. In Figure \ref{fig:variability} we plot the brightness of the progenitor candidate between 2013 and 2018, along with photometry of three nearby objects of similar brightness. The brightness of the progenitor candidate between 2013 and 2016 is consistent with the brightness we measure from the stack images. From the end of 2015 until the last epochs in 2018, the progenitor brightened by ${\sim}0.4$ and ${\sim}0.3$ magnitudes in the $g$ and $r$ bands, respectively. We believe this variability is astrophysical due to the smaller level of variability seen in the three comparison sources. To test the significance of this variability we perform a simple chi-squared test against a constant flux model. For both the $g$ and $r$ band light curves we find the reduced $\chi_r^2>3$, demonstrating that a constant flux model poorly describes the data. While this manuscript was in preparation, a similar study focusing on the variability of the progenitor candidate came to the same broad conclusions \citep{Niu2025}. Our independent analysis confirms their findings and further strengthens the evidence for variability. We are unable to search for periodicity due to the small number of observations, which makes determining the origin of the variability difficult. \citet{Niu2025}'s preferred explanation was stellar internal changes due to disturbances in the outer envelope caused by pulsations, episodic mass loss, and/or binary interactions. We discuss this further in Section \ref{sec:discussion}.
 
\subsection{Single-star models}\label{subsec:singlestar}
We investigate single-star models for the progenitor candidate by fitting the $griz$ DECam SED with two models: a simple blackbody function, and stellar SEDs from \citet{Pickles1998}. We exclude the VIRCAM upper limits from our analysis as they are not deep enough to meaningfully constrain any of the models. We use a forward modelling approach within the \verb|pysynphot| software package \citep{pysynphot} along with the DECam filter transmission curves\footnote{Obtained from the SVO Filter Profile Service\href{https://svo2.cab.inta-csic.es/theory/fps/}{https://svo2.cab.inta-csic.es/theory/fps/}} to determine $griz$ synthetic photometry for a given model. To correct for Galactic extinction we redden the SEDs for $E(B-V)=0.0089$~mag of extinction assuming a Milk Way extinction law with $R_V=3.1$ \citep{Cardelli1989}. We do not include host galaxy or circumstellar extinction in our modelling. For the blackbody fit the free parameters are the blackbody radius and temperature. The stellar flux library of \citet{Pickles1998} consists of 131 flux-calibrated spectra of stars of all main spectral types and luminosity classes at solar abundance, and were created by combining spectra from several sources. Each spectrum has an associated effective temperature and spectral type, and complete spectral coverage from 1150 to 10620~$\AA$. When fitting these spectra to our data the free parameters are a constant which sets the flux scale, and the temperature. 

We utilise the \verb|emcee| Python package to conduct a Markov chain Monte Carlo (MCMC) exploration of the model parameter space. Our likelihood function is a Gaussian and our priors are uniform in temperature, radius and $\log{F_\lambda}$. Due to the discrete nature of the \citet{Pickles1998} models, for a randomly sampled temperature we use the model with the nearest temperature to this value.  Prior to fitting we convert our observed magnitudes into absolute magnitudes using a distance modulus of $\mu=30.17\pm0.26$. We do not include the error on the distance modulus when fitting the data but we do include it in luminosity measurement errors. We report the median and the 16\textsuperscript{th} and 84\textsuperscript{th} percentiles (the $1\sigma$ credible interval) from our posterior distributions for all derived parameters henceforth. 

\begin{figure}
\centering
\includegraphics[width=\columnwidth]{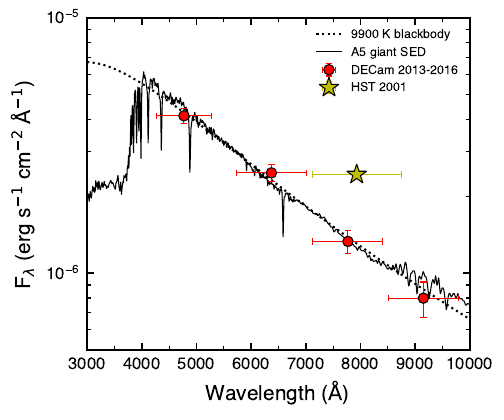}
\caption{DECam $griz$ SED of the progenitor candidate as determined from the stack images. Horizontal error bars indicate the wavelength range of each filter. The dotted line is the best-fit blackbody function with a temperature of 9880~K. The solid line is the best-fit stellar SED from \citet{Pickles1998} corresponding to an A5III supergiant with a temperature of 8453~K. The \HST{}/WFPC2 F814W flux measurement obtained in 2001 is a factor of ${\approx}1.9$ (0.7~mag) brighter than the $i$-band DECam measurement from 2013--2016. Fluxes are scaled to a distance of 10 parsecs.}
\label{fig:SED}
\end{figure}

For the blackbody fit we derive a temperature of $T=9900^{+1100}_{-800}$~K and a photosphere radius of $R=110^{+15}_{-14}~R_\odot$ with an associated luminosity of $\log(L/L_\odot)=4.99\pm0.16$. In contrast, the stellar SED which most closely matches the data has a lower temperature with $T=8453$~K and an A5III spectral type (Figure \ref{fig:SED}). The derived temperature from the MCMC fit is $T=8400^{+400}_{-500}$ with a larger radius of $R=124^{+18}_{-8}~R_\odot$ and a lower luminosity of $\log(L/L_\odot)=4.84\pm0.14$ compared to the blackbody fit. The discrepancy in temperature and luminosity between the two models arises because our photometric SED probes the Rayleigh-Jeans tail where differences between a blackbody and a realistic stellar atmosphere become more pronounced due to absorption features and UV line blanketing. We consider the stellar SED fit to be a more physically realistic representation of the data.  

\begin{figure}
\centering
\includegraphics[width=\columnwidth]{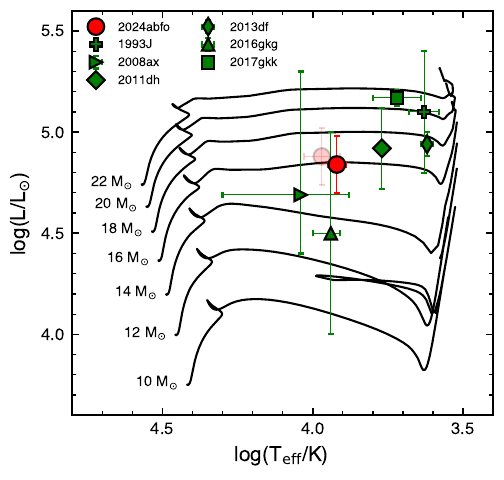}
\caption{Hertzprung-Russel diagram showing the location of SN~2024abfo as determined from the DECam stack images compared to six other Type IIb SNe with progenitor detections: 1993J \citep{Maund2004}, 2008ax \citep{Folatelli2015}, 2011dh \citep{Maund2011dh}, 2013df \citep{Vandyk2014}, 2016gkg \citep{Kilpatrick2022}, and 2017gkk \citep{Niu2024}. We show MIST evolutionary tracks for non-rotating, solar-metallicity stars. The faded red marker indicates the effect of $A_V=0.3$~mag of host-galaxy extinction on the SN~2024abfo progenitor candidate position.}
\label{fig:tracks}
\end{figure}

To estimate the initial mass of the progenitor candidate we compare the position of the progenitor in the Hertzsprung-Russel (H-R) diagram with single-star stellar evolution tracks from MESA Isochrones \& Stellar Tracks \citep[MIST;][]{Dotter2016,Choi2016}. The metallicity, rotation, and initial mass of the star are free parameters when generating the evolutionary tracks. We found that the effects of rotation (with $v_\mathrm{ZAMS}/v_\mathrm{crit}=0.4$) are negligible compared to tracks without rotation for stars in the mass range we are considering (see Figure 8 in \citet{Choi2016}), so we choose to show non-rotating models. The effect of metallicity is more pronounced, with lower metallicities resulting in hotter and more luminous tracks as a result of a lower Rosseland mean opacity (see Figure 9 in \citet{Choi2016}). Assuming solar metallicity, the track closest to the locus of the progenitor candidate in the H-R diagram is that of a star with an initial mass of 16 $M_\odot$. A different approach to estimate the initial mass is to use the end-point luminosity of a track since the temperature is not expected to be reflective of the star's initial mass, whereas the final luminosity depends more strongly on the size of the core which is connected to the initial mass, as pointed out by \citet{Maund2011dh}. Using this approach we estimate an initial mass of $13\pm2~M_\odot$. Including $A_V=0.3$ magnitudes of host galaxy extinction does not affect this result substantially, as demonstrated by the small change in position of the progenitor candidate in the H-R diagram in Figure \ref{fig:tracks}.
 
\subsection{Binary models}\label{subsec:binary}
Binary systems provide a natural mechanism for stripping the hydrogen-rich envelopes of Type IIb SN progenitors and have been invoked to explain Type IIb SNe including SN~1993J \citep{Maund2004,Fox2014} and 2011dh \citep{Folatelli2014,Maund2019}, among others. We explore binary models by comparing the location of the SN~2024abfo progenitor candidate in the H-R diagram with the endpoints of binary models from the Binary Population and Spectral Synthesis \citep[BPASS;][]{Eldridge2017,Stanway2018} set of models. In particular we make use of the Hoki Python package \citep{Stevance2020} to aid in querying the large number of BPASS models and parameters. 

\begin{figure*}
\centering
\includegraphics[width=\textwidth]{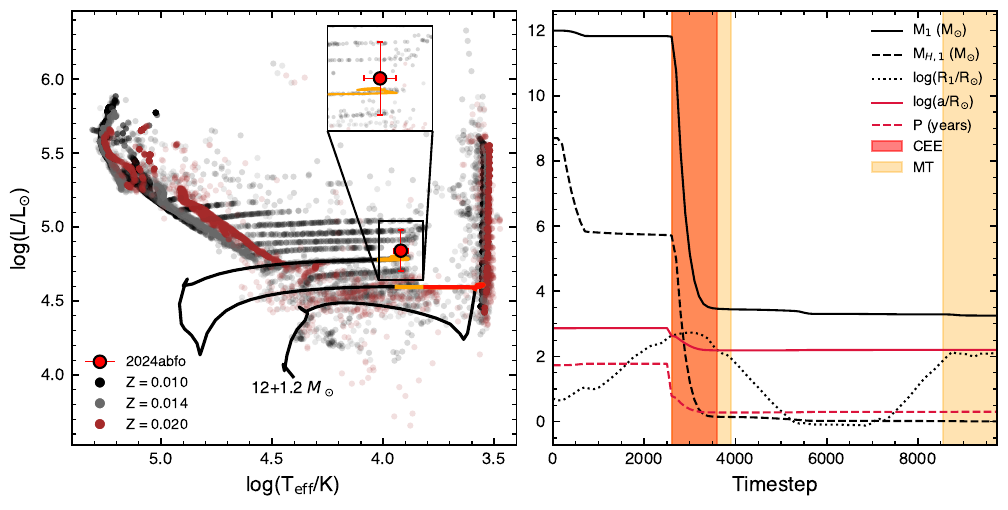}
\caption{Left: H-R diagram showing the endpoints of BPASS models for three different metallicities: $Z=0.010$, 0.014 and 0.020. The black line is the evolutionary track of a $12+1.2~M_\odot$ binary system with an initial period of 1.73 years. The red and orange segments of the track correspond to common envelope evolution (CEE) and mass transfer (MT) phases. The inset highlights the final evolution of the system prior to core-collapse of the primary. Right: Evolution of key parameters of the system, with the common envelope and mass transfer phases highlighted. The red solid and dotted lines denote the semimajor axis ($a$) and period ($P$). The black solid line and dashed lines denote the mass of the primary star ($M_1$) and its hydrogen envelope ($M_{H,1}$). Time steps do not correspond to equal time intervals. }
\label{fig:BPASS}
\end{figure*}

BPASS models are available for 13 different metallicities. Individual models are parametrised by a further three parameters: the two masses of the binary components ($M_\mathrm{1}$,$M_\mathrm{2}$), and their orbital period at formation ($P_\mathrm{ZAMS}$). The metallicity of the nuclear star cluster in NGC~1493 is solar (with $Z=0.02$; Section \ref{sec:discovery}). Many galaxies show a decreasing metallicity gradient with increasing radius from the centre: the Milky Way has a gradient of $-0.07$~dex/kpc for light elements including C, O, Mg and Si \citep{Rolleston2000}. We therefore choose to consider solar ($Z=0.020$) and slightly sub-solar ($Z=0.014$ and $0.010$) metallicities for the location of SN~2024abfo. We narrow down the number of models by only considering those where the primary is likely to undergo core-collapse: we set a lower mass limit on the initial mass of the primary at $7~M_\odot$ and constrain the CO and ONe core masses to be greater than 1.38 and 0.1 $M_\odot$ at death\footnote{These are used in the Hoki tutorial \href{https://heloises.github.io/hoki/MIAPBP_EvE_tutorial.html}{Searching for specific star systems in BPASS using EvE}, a very useful resource for working with BPASS models.}. Figure \ref{fig:BPASS} compares the endpoints of the BPASS models in the H-R diagram for the three metallicities considered. The effect of metallicity is pronounced: only a single model at solar metallicity is consistent with the position of 2024abfo, whereas this increases to 5 models for $Z=0.014$ and 66 models for $Z=0.010$. A sub-solar metallicity of $Z=0.010$ is therefore preferred, which would also be consistent with an expected metallicity gradient relative to the centre of the galaxy. 

The probability of a specific model is related to the number of occurrences of the system in a $10^6~M_\odot$ population assuming the default BPASS initial mass function \citep{Stevance2021}. Among the 66 sub-solar metallicity models consistent with the progenitor position, we calculate the number of systems grouped by primary mass. The most probable primary mass is $12~M_\odot$ (27\%) followed by $11~M_\odot$ (21\%). The remainder of systems have primaries with masses ranging from 9--15~$M_\odot$. The most probable $12~M_\odot$ system has a secondary companion with a mass of $1.2~M_\odot$ and a binary period of 1.73 years at the time of formation. We show the evolution of this system in the H-R diagram in Figure \ref{fig:BPASS}, along with the evolution of selected parameters of the system. The hydrogen mass of the primary decreases during the main sequence as hydrogen is fused into helium, while the stellar radius gradually increases. When the primary expands sufficiently to overflow its Roche lobe, a phase of unstable mass transfer begins. During this common envelope evolution (CEE), the hydrogen-rich envelope of the primary is rapidly ejected from the system leading to a significant decrease in the mass of the primary and its hydrogen envelope. The orbital separation and binary period decrease due to angular momentum loss. After core helium burning ends the primary undergoes helium shell burning which causes its radius to increase again, which leads to a final period of mass transfer lasting $\approx4000$ years prior to collapse of the ONe core. At the time of explosion the primary has a mass of $3.25~M_\odot$ with a hydrogen envelope mass of ${\sim}0.01~M_\odot$ and a radius of $125~R_\odot$, almost identical to the value of $R=124^{+18}_{-8}~R_\odot$ derived from single star models in Section \ref{subsec:singlestar} above. Mass transfer and shell burning instabilities can lead to small back and forth movements within the H-R diagram (see inset in Figure \ref{fig:BPASS}), which may explain the observed variability seen in the progenitor candidate (Section \ref{subsec:variability}) in the final years prior to explosion. We caution, however, that the model shown here is one of many whose endpoint in the H-R diagram is consistent with the progenitor candidate. 

\section{The Supernova}\label{sec:supernova}
\subsection{Light curves}
\begin{figure*}
\centering
\includegraphics[width=0.85\textwidth]{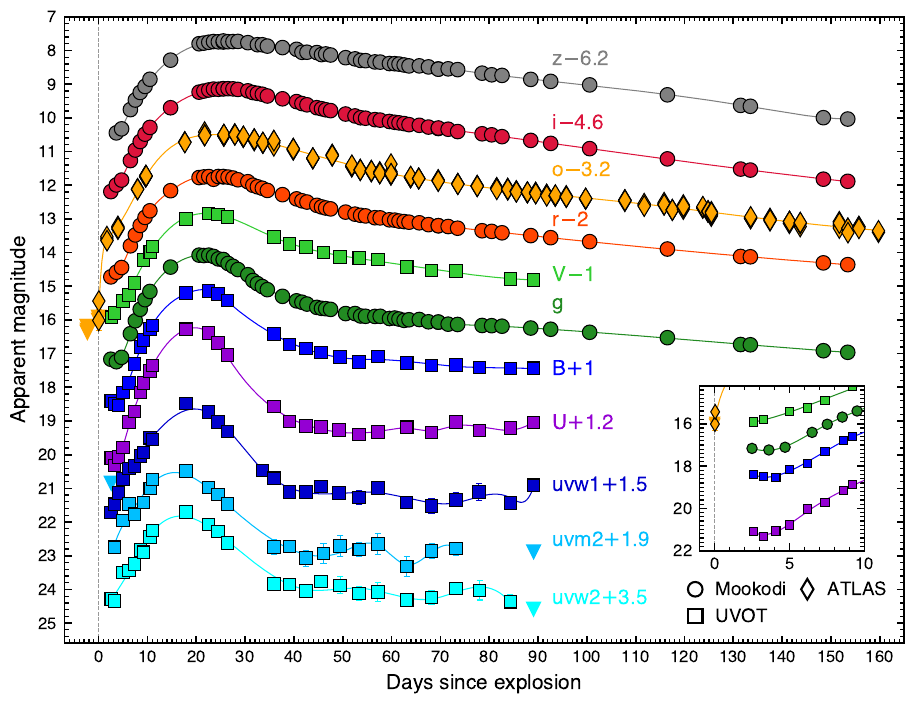}
\caption{UV/optical light curves for SN 2024abfo with smooth cubic spline fits to each light curve. Error bars are in most cases smaller than the symbols. All magnitudes are in the AB system. The vertical dotted line denotes the explosion epoch (MJD 60628.28) derived from the intra-night ATLAS discovery. Upper limits are shown with upside-down triangles and are at the $3\sigma$ level. The inset highlights the short-lived shock cooling phase visible in the $g$, $B$ and $U$ bands.}
\label{fig:all_LC}
\end{figure*}

We show the photometric evolution of SN 2024abfo as detected by Mookodi, ATLAS, and \Swift{}/UVOT in Figure \ref{fig:all_LC}. The explosion date is constrained by the intra-night ATLAS discovery (Section \ref{sec:discovery}). In the $g$, $B$ and $U$ bands there is tentative evidence for a short-lived shock cooling phase prior to 5 days, which we discuss further in Section \ref{subsec:shock}. To characterize the light curves further, we fit cubic spline fits to each band to determine the peak time and brightness and the decline over the first 15 days post-peak (see Table \ref{tab:LC_peaks}). Like many stripped-envelope SNe, the peak is reached earlier in the bluer bands compared to the redder bands, with the $U$ peak occuring 5.2 days before the $r$ peak. 

\begin{figure*}
\centering
\begin{subfigure}[b]{0.45\textwidth}
    \includegraphics[]{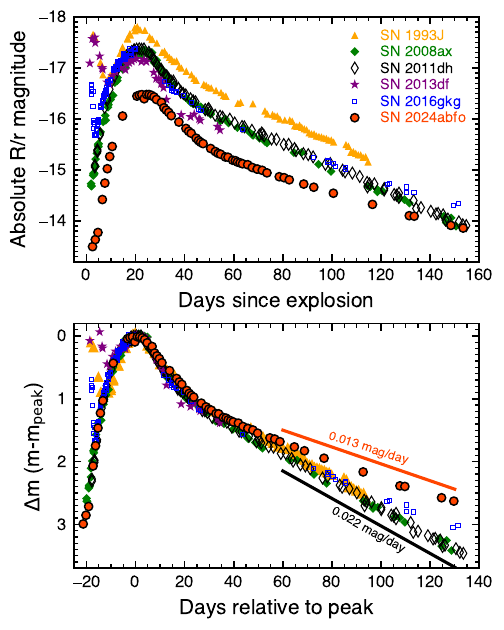}
    \caption*{} 
\end{subfigure}
\hfill
\begin{subfigure}[b]{0.45\textwidth}
    \includegraphics[]{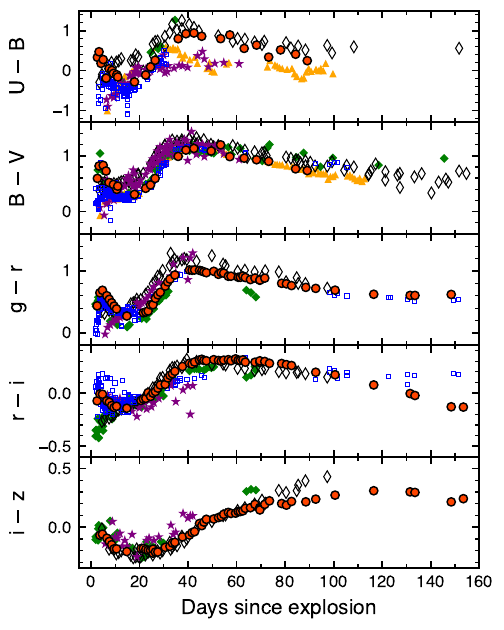}
    \caption*{} 
\end{subfigure}
\vspace{-1em}
\caption{
Left: Absolute $R/r$-band light curves of Type IIb supernovae with progenitor detections (top) and the same light curves shifted to a common peak time and magnitude (bottom). We highlight the decline rate between 60 and 140 days for \SN{} and SN~2011dh. Right: Colour curves of \SN{} compared to the same sample. We include data from \citet{Szalai2016} for the colour curves.
}
\label{fig:comparison}
\end{figure*}

In the top left panel of Figure \ref{fig:comparison} we compare the $R/r$-band absolute magnitude light curves of Type IIb supernovae which have progenitor detections, excluding SN~2017gkk, which lacks published photometric data. We use published light curves from \citet{Richmond1996} for SN~1993J; \citet{Pastorello2008} and \citet{Tsevtkov2009} for SN~2008ax; \citet{Ergon2015} for SN~2011dh; \citet{Morales2014} for SN~2013df; and \citet{Tartaglia2017} and \citet{Kilpatrick2022} for SN2016gkg. Each light curve is corrected for extinction and converted to absolute magnitudes using values reported in the respective sources. Among these events, SN~2024abfo is the least luminous, peaking at an absolute magnitude ${\sim}0.5$ magnitudes fainter than SN~2013df, the next faintest object. Three of these SNe---1993J, 2013df and 2016gkg---show prominent shock-cooling phases in their early light curves. SN~2024abfo is therefore more similar to SNe 2008ax and 2011dh, though SN~2011dh did show a brief shock-cooling phase at early times in the $g$ and $r$ bands \citep{Arcavi2011}. 

The lower left panel of Figure \ref{fig:comparison} demonstrates that the shape of the $R/r$-band light curves during the main peak is broadly similar across these events, aligning from the earliest times (if one excludes the shock-cooling SNe) up until ${\sim}20$ days post-peak. Thereafter, SN~2024abfo appears to deviate from the rest of the sample by following a shallower decay: its decline rate between 60 and 140 days is 0.013 mag/day compared to 0.021 mag/day for SN~2011dh. Both rates are steeper than the expected decline of 0.0098 mag/day from complete \textsuperscript{56}Co energy deposition, likely due to $\gamma$-ray leakage. Longer-term monitoring will be required to determine whether the shallower decline in \SN{} persists.

\begin{table*}
\caption{Light curve parameters derived from cubic spline fits.}
\label{tab:LC_peaks}
\centering
\begin{tabular}{lcccc}
\hline\hline
Band & \makecell{Peak time \\ (days)\tablefootmark{a}} & \makecell{Peak apparent \\magnitude} & \makecell{Peak absolute \\magnitude\tablefootmark{b}} & \makecell{$\Delta m_{15}$ \\(mag)\tablefootmark{c}} \\
\hline
$uvw2$ & $17.1\pm0.5$ & $18.29\pm0.08$ & $-11.88\pm0.08$ & $1.54\pm0.19$ \\
$uvm2$ & $16.0\pm0.4$ & $18.58\pm0.10$ & $-11.59\pm0.10$ & $1.64\pm0.23$ \\
$uvw1$ & $19.4\pm0.7$ & $17.11\pm0.05$ & $-13.06\pm0.05$ & $1.99\pm0.17$ \\
$U$ & $18.9\pm0.4$ & $15.05\pm0.03$ & $-15.12\pm0.03$ & $2.07\pm0.12$ \\
$B$ & $20.7\pm0.7$ & $14.12\pm0.03$ & $-16.05\pm0.03$ & $1.29\pm0.08$ \\
$g$ & $21.6\pm2.0$ & $14.08\pm0.02$ & $-16.09\pm0.02$ & $1.17\pm0.23$ \\
$V$ & $22.1\pm0.5$ & $13.85\pm0.02$ & $-16.32\pm0.02$ & $0.76\pm0.04$ \\
$r$ & $24.7\pm2.4$ & $13.75\pm0.07$ & $-16.42\pm0.07$ & $0.63\pm0.15$ \\
$o$ & $24.3\pm7.4$ & $13.70\pm0.01$ & $-16.47\pm0.01$ & $0.43\pm0.39$ \\
$i$ & $26.1\pm2.3$ & $13.74\pm0.06$ & $-16.43\pm0.06$ & $0.47\pm0.18$ \\
$z$ & $26.0\pm2.8$ & $13.93\pm0.17$ & $-16.24\pm0.17$ & $0.35\pm0.26$ \\
\hline
\end{tabular}
\tablefoot{\\
\tablefoottext{a}{Observer frame time with respect to MJD 60628.28.} \\
\tablefoottext{b}{Calculated assuming a distance modulus of $\mu=30.17\pm0.26$.} \\
\tablefoottext{c}{Decline within 15 days of peak.}
}
\end{table*}

\subsection{Color evolution}
In the right panel of Figure \ref{fig:comparison} we show the $U-B$, $B-V$, $g-r$, $r-i$ and $i-z$ colour curves for \SN{} compared to the sample of IIb SNe with progenitor detections. We correct the photometry for Galactic extinction before computing each colour. \SN{} aligns closely with the majority of the colour curves, though it appears to show a red excess at early times in $U-B$. The supernovae with strong shock cooling phases (e.g. 1993J, 2013df and 2016gkg) are bluer at this time, which suggests that \SN{} is powered by radioactive heating rather than shock cooling during this phase. The $B-V$ colour aligns closely with SN~2008ax at early times with both objects appearing redder than the rest of the sample, which is consistent with the fact that both events lacked a prominent shock cooling phase.  The lack of a redward deviation in the $U-B$ and $B-V$ colour evolution suggests that the host galaxy extinction is not significant

A noticeable deviation occurs in the $r-i$ colour after ${\sim}100$ days, where \SN{} appears to evolve blueward compared to SN~2016gkg, the only other SN with coverage at these times. Although the deviation is modest (${\sim}0.3$ magnitudes at 150 days), it may reflect the shallower decline seen in the $r$-band light curve compared to the rest of the sample (bottom left panel in Figure \ref{fig:comparison}). 

\subsection{Bolometric light curve}
\begin{figure}
\centering
\includegraphics[width=\columnwidth]{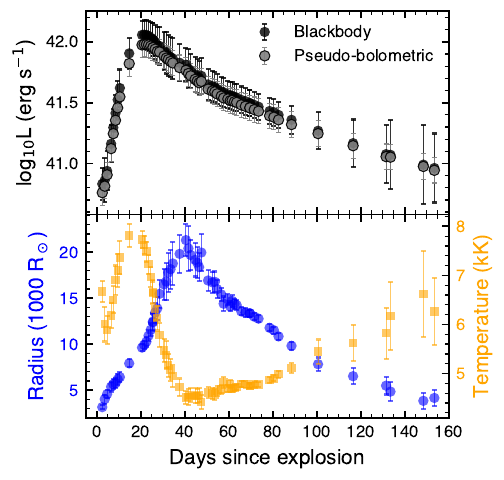}
\caption{Top: Pseudobolometric UV/optical light curve compared to the bolometric luminosity derived from blackbody fits to each SED. Errorbars include the uncertainty in the distance modulus. Bottom: Temperature and radius evolution as derived from the blackbody fits. \label{fig:Lbol}}
\end{figure}

To construct a bolometric light curve we interpolate our multi-band light curves to the times of the $g$-band epochs using our spline fits. We then fit the SED at each epoch following the approach described in Section \ref{subsec:singlestar} for the progenitor SED fitting, whereby we use synthetic photometry from a reddened blackbody spectrum to determine the best-fit temperature and radius. At early times the effect of UV line blanketing leads to a suppression of the flux in the \Swift{}/UVOT ultra-violet bands, so we choose to exclude these bands at all times when fitting. After 90 days the \Swift{}/UVOT coverage ends so we fit blackbodies only to the $griz$ photometry. At each epoch we also calculate a pseudo-bolometric luminosity by integrating the SED in flux density units ($F_{\lambda}$) over the observed wavelength range, after correcting the observed photometry for Galactic extinction. At wavelengths redder than those considered here, the contribution to the blackbody luminosity is considerable, ranging from 18\% for the hottest temperatures to 44\% at the coolest. We therefore include this contribution in the pseudo-bolometric luminosity. After 90 days we also include the blackbody contribution at wavelengths blueward of the $g$-band. Figure \ref{fig:Lbol} compares the pseudo-bolometric luminosity with the luminosity assuming a perfect blackbody, along with the temperature and radius evolution. Between 2 and 5 days the temperature shows a short and steep drop, which can be interpreted as the tail-end of a short-lived shock-cooling phase. Thereafter the temperature rises to a maximum of $T{\sim}7800$~K at ${\sim}17$ days post-explosion before declining. The radius of the photosphere increases monotonically until reaching a peak of $R{\sim}21000~R_\odot$ at 40 days, which is also when the temperature reaches a minimum. Thereafter the radius and temperature decline and rise, respectively, until our photometric monitoring ends at 153.5 days. 

\begin{figure}
\centering
\includegraphics[width=\columnwidth]{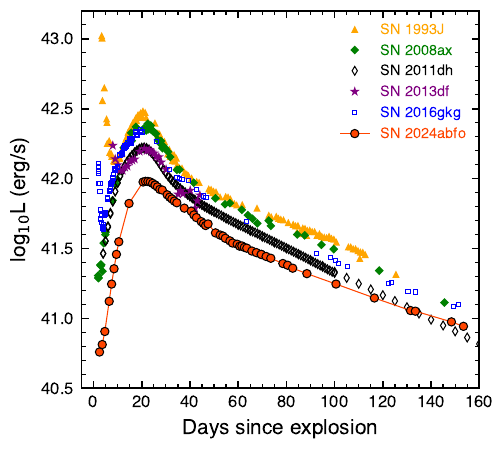}
\caption{Comparison of bolometric light curves for Type IIb SNe with progenitor detections.\label{fig:Lbol_comparison}}
\end{figure}

In Figure \ref{fig:Lbol_comparison} we compare the bolometric light curves of the Type IIb SNe with progenitor detections. For SNe 1993J and 2011dh we obtain their bolometric light curves directly from the literature \citep{Ergon2015,Lewis1994}, while for the remaining SNe we use the equations in \citet{Lyman2014} to compute bolometric corrections from $B-I$ and $g-r$ colours. Similar to the $R/r$-band absolute magnitude light curves, \SN{} is clearly the least luminous object in the sample. When all of the objects are shifted to the same peak time and luminosity (see inset of Figure \ref{fig:Lbol_comparison}), it is apparent that the late-time decay rate is shallower than the comparison sample, as was also seen in the $R/r$-band comparison.  

\subsection{Shock cooling emission and light curve modelling} \label{subsec:shock}
The presence of an early shock-cooling peak can be used to derive constraints on the envelope mass and its radius \citep[e.g. SN~2016gkg;][]{Arcavi2017} using a variety of analytical models in the literature \citep[e.g.][]{Nakar2010,Rabinak2011,Nakar2014,Piro2015,Sapir2017,Morag2023}. The bolometric light curve of \SN{} does not show a prominent shock-cooling peak but it does show a very weak and short-lived declining phase in the $g$, $B$ and $U$-band light curves (Figure \ref{fig:all_LC}) which may be the tail-end of shock-cooling emission. We would like to apply these models to our data in order to derive constraints on the mass and radius of the progenitor envelope, however, \citet{Sapir2017} made clear that their analytic expressions are only valid while the opacity in the stellar envelope is that of a fully ionised gas with temperatures $T>0.7$~eV (8120~K), since at lower temperatures recombination leads to a decrease in the opacity and the photosphere recedes into deeper ejecta where there is a stronger dependence on the envelope structure. In this regime, detailed radiation transfer models are required. The blackbody temperature derived from our light curves is always less than 8120~K (Figure \ref{fig:Lbol}), which precludes us from using the analytical models of \citet{Sapir2017} to constrain the radius of the progenitor's envelope. 

Instead, we choose to use the two-component semi-analytic model of \citet{Nagy2016} to model the main radioactive-powered peak and place constraints on the radius and mass of the extended outer envelope of the progenitor due to a short-lived or weak shock-cooling phase. The model of \citet{Nagy2016} includes the effects of recombination as first presented in \citet{Arnett1989}, and generalises the model presented by \citet{Nagy2014}. The model assumes homologously expanding spherical ejecta with two separate components: a constant-density inner "core" which is denser, more massive and He-rich; and an outer "envelope" with a power-law density structure which is extended, low-mass and H-rich. This structure is intended to reflect the make-up of red or yellow supergiants which have an inflated outer envelope surrounding a compact core. Such a configuration was used by \citet{Bersten2012} and \citet{Kumar2013} to model the Type IIb SNe 2011dh and 2011fu. The model solves the photon diffusion equation in the ejecta and takes into account the energy released due to recombination and radioactive heating. Since the photon diffusion time is much lower in the outer shell compared to the inner core, the diffusion equation can be solved separately for each component \citep{Kumar2013}, and the total bolometric luminosity is therefore the sum of the luminosity from the core and envelope components. A constant Thompson-scattering opacity is assumed for each component (though see \citet{Nagy2016} for a more detailed discussion). 

\begin{figure*}
\centering
\includegraphics[width=\textwidth]{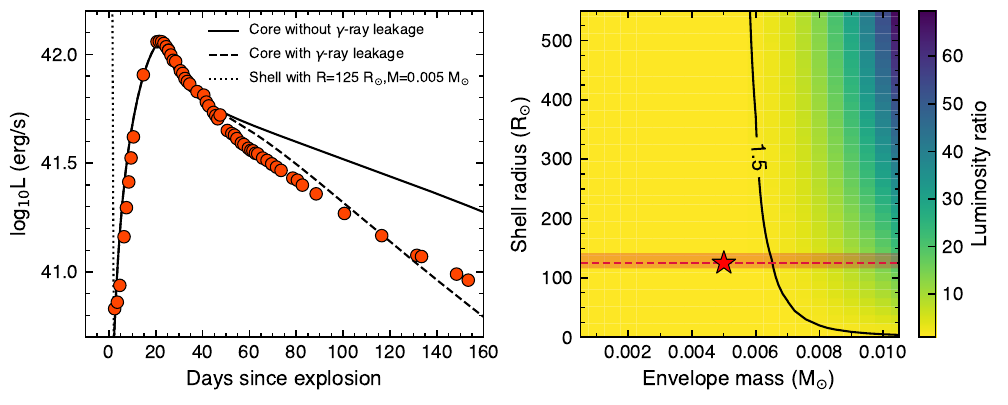}
\caption{Left: \SN{} bolometric light curve along with our best-fit core model of \citet{Nagy2016} with and without $\gamma$-ray leakage (dashed and solid lines, respectively). The dotted line is the shell component for an envelope with a radius of $125~R_\odot$ and a mass of $0.005~M_\odot$. Right: Constrained parameter space for varying shell radius and envelope mass. The colour bar denotes the ratio of the total (core+shell) model luminosity with respect to the measured luminosity at 2.5 days. The contour line indicates where the model luminosity is 50\% (0.44~magnitudes) larger than the measured luminosity. We regard all models to the left of this line to be consistent with our observations. The red dashed line and shaded region correspond to the radius we measured for the progenitor from single star models (Section \ref{subsec:singlestar}). The corresponding shell model light curve for the parameters at the position of the red star are shown in the left panel. \label{fig:bol_model}}
\end{figure*}

Due to the large number of free parameters and the strong correlations between them, we do not adopt a formal fitting procedure to determine the best-fit parameters for the core component. Instead we manually adjust the parameters until a satisfactory fit is achieved, as done by \citet{Nagy2016} and \citet{Szalai2016}. We choose to fix certain parameters at the same values as those used by \citet{Szalai2016} for the Type IIb SN~2013df. In particular, we set the power-law density profile index of the envelope component to $n=2$, consistent with a stellar wind, and the recombination temperature in the core to a temperature of 10 000~K. For the opacities in the core and envelope we adopt their values of $\kappa=0.2$ and 0.4~cm$^2$~g$^{-1}$, which are consistent with the average opacities produced by the SNEC hydrodynamic code for Type IIb SNe \citep{Morozova2015,Nagy2016}. We find that the initial radius of the core is constrained to $R_0\lesssim10^{11}$~cm or $1.44~R_\odot$, the mass of synthesised \textsuperscript{56}Ni is $M_\mathrm{Ni}=0.055~M_\odot$, the ejecta mass is $M_\mathrm{ej}=2.55~M_\odot$, and the kinetic energy in the ejecta is $E_\mathrm{k}=1.95\times10^{51}$~erg. As pointed out by \citet{Szalai2016}, only the degenerate combinations of $M_\mathrm{ej}\kappa$ and $E_\mathrm{k}\kappa$ can be constrained by the observations, so assuming a smaller (larger) value for the opacity would result in larger (smaller) values for the ejecta mass and kinetic energy. They also point out that the values of $M_\mathrm{ej}$ and $E_\mathrm{k}$ derived from modelling the peak versus the tail of the light curve can result in very different results. We do not model the late time light curve as we consider it beyond the scope of this work. Our model light curve (Figure \ref{fig:bol_model}) shows a reasonable fit to the bolometric light curve until ${\sim}40$ days. After this time the observed light curve decays faster than the model, which may indicate that there is $\gamma$-ray leakage. The model of \citet{Nagy2016} includes the ability to account for $\gamma$-ray leakage. We are able to generate a better fit in terms of the luminosity at later times, but this model cannot capture the decay slope, as shown in Figure \ref{fig:bol_model}. 

To constrain the radius and mass of the outer envelope we generate model light curves for the shell component using a grid of models varying in envelope mass and radius. We fix the kinetic and thermal energies in the shell at the values used in \citet{Szalai2016}. The total bolometric luminosity is the sum of the core and shell components. We calculate the ratio of the model luminosity with respect to the measured luminosity at the time of our first data point at 2.5 days and use this as a means of determining which model parameters are consistent with our observed bolometric light curve. We consider a model to be consistent with our observations if the model flux at 2.5 days is a factor of 1.5 (50\%) or less than the measured flux at this time. The right panel of Figure \ref{fig:bol_model} demonstrates that larger radii result in longer-lived and more luminous shock-cooling peaks when the envelope mass is ${\sim}0.01M_\odot$. As the envelope mass decreases, however, we are unable to constrain the radius of the stellar envelope, and all models with an envelope mass $\lesssim0.006~M_\odot$ are essentially consistent with the data, regardless of the radius. The models which are consistent with our observations have a rapid and luminous shock cooling peak which decays within 2.5 days of the explosion, as shown by the shell model in the left panel of Figure \ref{fig:bol_model}. If we fix the radius of the envelope to the radius we found from our progenitor analysis (Sections \ref{subsec:singlestar} and \ref{subsec:binary}) of $R\approx125~R_\odot$, we find that all envelopes with a mass of $<0.0065~M_\odot$ are still consistent with our bolometric light curve. This value is close to the envelope mass of ${\sim}0.01~M_\odot$ we found from the most probably BPASS model, although this particular model (with $R=125~R_\odot$) would be inconsistent with our observations. According to \citet{Dessart2011} and \citet{Bersten2012}, the presence or absence of a shock cooling peak depends on the mass of the H-rich envelope. The low envelope mass of $<0.0065M_\odot$ required for \SN{} is therefore consistent with the very weak shock cooling emission observed in \SN{}. 

\subsection{Spectra}\label{subsec:spectra}
\begin{figure*}[p]
\centering
\includegraphics[width=\textwidth]{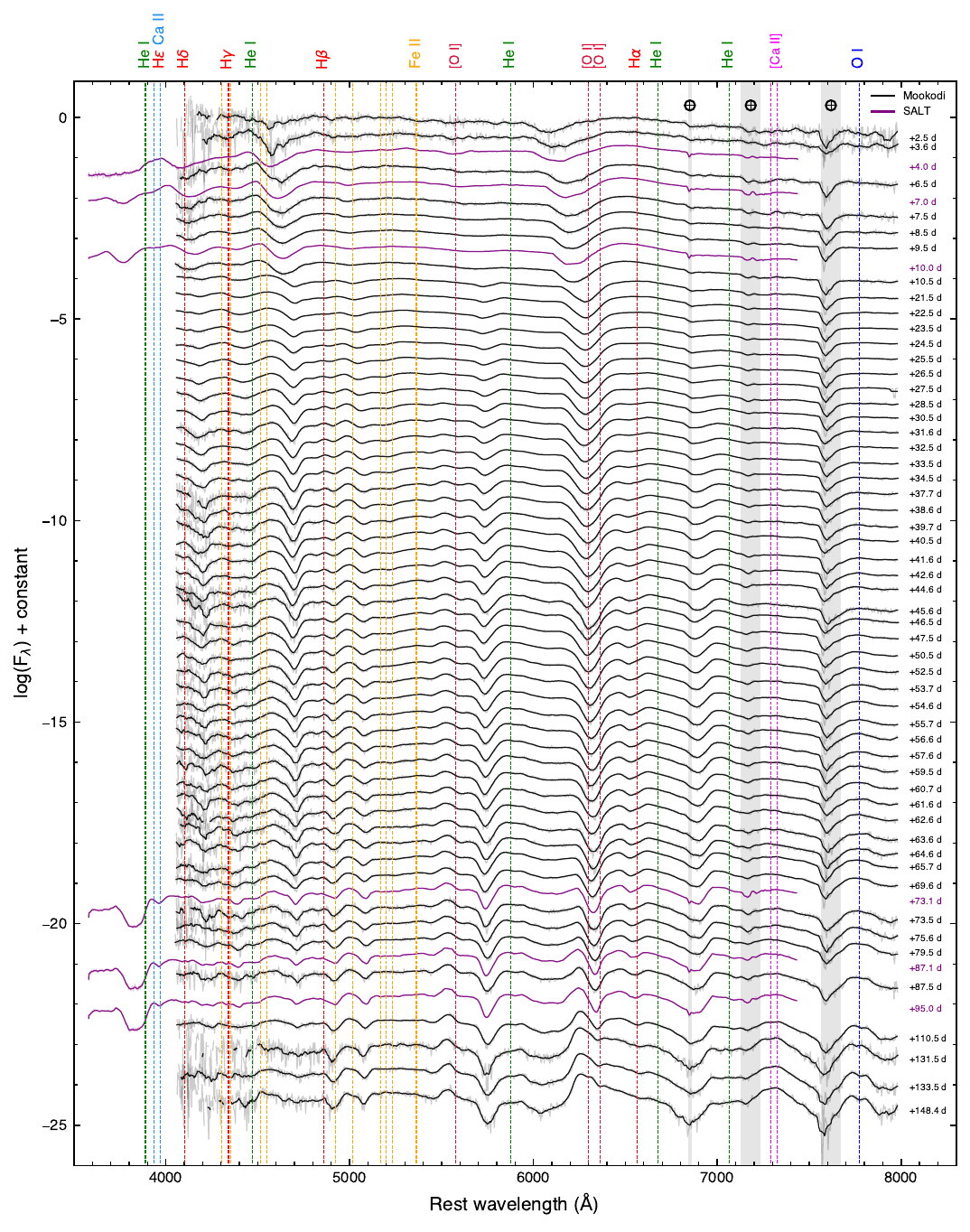}
\caption{Spectral series of \SN{} obtained with Lesedi/Mookodi and SALT. The rest wavelengths of prominent atomic species are indicated via vertical dashed lines. Vertical grey regions denote telluric absorption. Noisy spectra have been smoothed with a Savitzky-Golay filter, as indicated by the black solid lines. }
\label{fig:all_spec}
\end{figure*}

\begin{figure}
\centering
\includegraphics[width=\columnwidth]{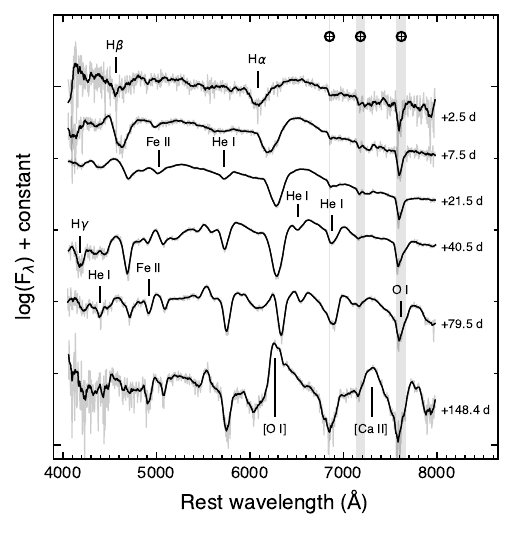}
\caption{Lesedi/Mookodi spectra of \SN{} at five epochs ranging from our first spectrum at 2.5 days to our last spectrum at 148.4 days post-explosion, with the most prominent features highlighted. \label{fig:line_IDs}}
\end{figure}

In Figure \ref{fig:all_spec} we present our 59 spectra ranging from 2.5 days to 148.4 days post-explosion. Our very first spectrum shows broad P-Cygni H$\alpha$ and H$\beta$ blue-shifted absorption lines, which led to the initial classification of \SN{} as a Type II SN but with no sub-type given \citep{2024TNSAN.342}. In Figure \ref{fig:line_IDs} we show spectra at six representative epochs with the most obvious spectral features marked. Our spectrum at 21.5 days shows \ion{He}{i}~$\lambda5876$ in absorption, firmly establishing the subtype as IIb. This feature was in fact visible earlier but its identification was unclear due to it exhibiting a broad and shallow absorption profile, as seen in the 7.5 day spectrum. At ${\sim}25$ days the \ion{He}{i}~$\lambda6678$ absorption feature starts becoming more prominent, and by 40.5 days there is an obvious dip in the spectrum near the broad H$\alpha$ emission peak (Figure \ref{fig:line_IDs}). Throughout the spectral evolution, \ion{Fe}{ii} absorption lines are visible between 4700 and 5300 $\AA$. By the time of our final spectrum at 148.4 days, \SN{} is clearly in the nebular phase with strong emission lines of [\ion{O}{i}] and [\ion{Ca}{ii}] visible. 

\begin{figure}
\centering
\includegraphics[width=\columnwidth]{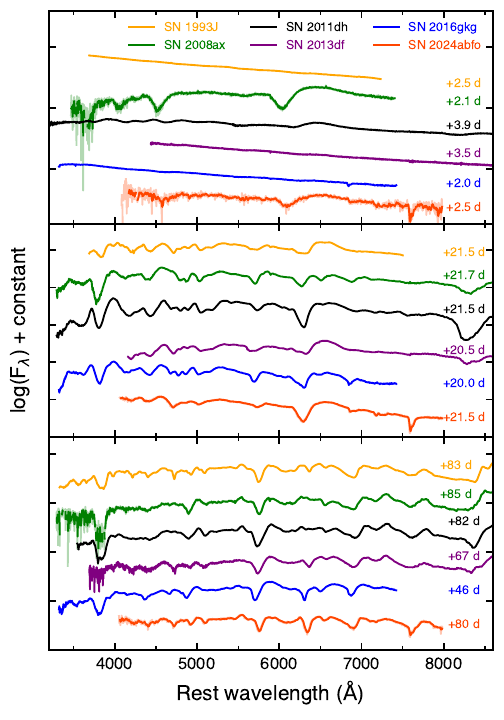}
\caption{Spectral comparison at three epochs of Type IIb SNe with direct progenitor detections. The top panel presents the earliest spectrum available for each object. The middle panel presents spectra at optical peak which occurs around 21 days post-explosion. The bottom panel presents spectra during the late photospheric/eraly nebular phase around 80 days post-explosion. \label{fig:spec_comparison}}
\end{figure}

\begin{figure}
\centering
\includegraphics[width=\columnwidth]{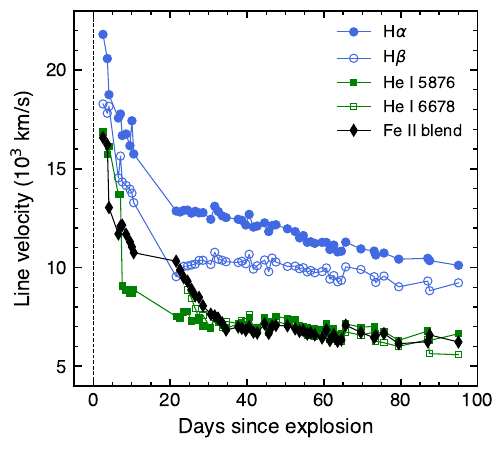}
\caption{Line velocities for SN{} as measured from the absorption minimum for each line. For the \ion{Fe}{ii} blend we use the \ion{Fe}{ii}~$\lambda 5198$ line as the reference wavelength. We do not show errorbars as the uncertainty is dominated by the spectral resolution for Mookodi. The typical uncertainty in the velocity is $\Delta v\approx1700$ km/s at 6000~$\AA$. \label{fig:line_velocities}}
\end{figure}

In Figure \ref{fig:spec_comparison} we compare the spectral evolution of \SN{} with Type IIb SNe having direct progenitor detections. We obtain archival spectra from the Weizmann Interactive Supernova Data Repository \citep[WISeREP;][]{WISEREP}. The top panel shows the earliest spectra for each of these SNe, ranging from 2 to 4 days post-explosion. There are clear spectral differences between the SNe at these early stages. The spectra of SNe 1993J, 2013df and 2016gkg are mostly blue and featureless and coincide with a prominent shock-cooling phase seen in their optical light curves (Figure \ref{fig:comparison}). SN 2011dh shows broad and shallow Balmer absorption lines, but with an overall smooth continuum. \SN{} is most similar to SN~2008ax, with both objects showing prominent Balmer absorption lines in their earliest spectra. The presence of strong shock-cooling emission in the early optical light curves results in hotter, smoother spectra, as the hot, ionized ejecta have a higher continuum opacity relative to the line opacity. As the ejecta expand and cool the photosphere recedes and line absorption becomes stronger. SN~2011dh did not show a prominent shock-cooling phase in its bolometric light curve but it did exhibit a short-lived shock cooling signature in the $g$ and $r$ bands \citep{Arcavi2011}, which may explain the intermediate nature of its early spectrum---smooth but with broad absorption features. By the time of the optical peak at ${\sim}21$ days, the spectra of all the SNe appear quite similar but with small differences. Helium lines have appeared, but the \ion{He}{i}~$\lambda6678$ absorption feature is weaker in SN~2013df and \SN{} compared to the rest of the sample. The strength of the \ion{He}{i}~$\lambda5876$ line also appears weaker in \SN{} compared to the rest of the sample, while H$\beta$ appears stronger. We show the velocity evolution of H$\alpha$, H$\beta$, \ion{He}{i}~$\lambda5876$, \ion{He}{i}~$\lambda6678$ and the \ion{Fe}{ii} blend centered at $5198~\AA$ in Figure \ref{fig:line_velocities}. The velocity in all lines drops rapidly in the first 10 days and reaches a plateau around 20 days, although the \ion{Fe}{ii} blend and \ion{He}{i}~$\lambda6678$ lines appear to show a steady linear decline from 20 to 35 days prior to flattening. Thereafter all lines decrease slowly by ${\sim}1000$~km/s until 100 days.  

A defining characteristic of the spectral evolution of \SN{} is the continued strength of H$\alpha$ over time. This is in contrast to the other Type IIb SNe with progenitor detections. At ${\sim}80$~days the H$\alpha$ line is barely visible in SNe 1993J, 2008ax and 2011dh at similar phases. For SNe 2013df and 2016gkg H$\alpha$ is visible, but the spectra we are able to obtain which are closest in phase are at 67 and 46 days post-explosion, when H$\alpha$ is expected to be stronger. We measure the pseudo-equivalent width \citep[pEW;][]{Gutierrez2017} of the H$\alpha$ absorption feature in each of these spectra and find that the pEW of H$\alpha$ is twice as large (or more) compared to SNe 1993J, 2008ax and 2011dh at the same epoch, and larger than the remaining SNe despite the earlier phases of SNe 2013df and 2016gkg (Table \ref{tab:pEW}). \citet{Liu2016} compared the pEW evolution of a sample of Type IIb and Ib SNe and showed that the strength of H$\alpha$ can be used to distinguish Type IIb from Ib SNe at all epochs. There are very few objects in their sample with measured pEW values at 80 days, however, the measured pEW in \SN{} is stronger than all other objects except for SN 2011ei \citep{Milisavljevic2013}. Interestingly, SN~2011ei was one of the least luminous SNe IIb or Ib observed at the time with a peak absolute magnitude of $M_V\approx-16$, which is only 0.3 magnitudes fainter than the peak $V$-band magnitude of \SN{}. The persistence of strong hydrogen absorption may suggest a higher abundance of hydrogen within the inner ejecta or a lower ionisation state compared to SNe in the comparison sample, however, detailed non-local thermodynamic equilibrium (LTE) modeling is required to determine the origin of the persistent hydrogen absorption.

\begin{table}
\caption{Pseudo-equivalent width (pEW) of H$\alpha$}
\label{tab:pEW}
\centering
\begin{tabular}{lcc}
\hline\hline
Supernova & Phase (days) & pEW ($\mathrm{\AA}$) \\
\hline
1993J       & 83 & $35.1 \pm 1.0$ \\
2008ax      & 85 & $15.2 \pm 0.3$ \\
2011dh      & 82 & $23.0 \pm 0.9$ \\
2013df      & 67 & $51.7 \pm 0.2$ \\
2016gkg     & 46 & $61.4 \pm 0.4$ \\
\textbf{2024abfo} & 80 & $77.7 \pm 1.6$ \\
\hline
\end{tabular}
\end{table}

\section{Discussion}\label{sec:discussion}

Before \SN{}, a total of six Type IIb SNe had direct progenitor detections. \SN{} is now the seventh Type IIb SN with a direct progenitor detection, although the association will need to be confirmed via late-time imaging to confirm its disappearance. An important question is whether the progenitor arose from a single star or binary system. 

From our modelling of the progenitor SED (Section \ref{subsec:singlestar}) the progenitor candidate appears to be an A-type supergiant with a radius of ${\approx}125~R_\odot$. The position of the progenitor in the H-R diagram sits squarely in the yellow void between red supergiants (RSG) and Wolf-Rayet stars. Considering only single stars, \citet{Groh2013} predicted that the progenitors of IIL and IIb SNe are yellow hyper giants (YHGs) or luminous blue variables (LBVs) for rotating models, with initial masses in the range 16.8--19~$M_\odot$ and 19--25~$M_\odot$, respectively. For non-rotating models the progenitors are either RSGs with masses in the range 19--23~$M_\odot$ or LBVs with masses in the range 23--32~$M_\odot$. The radii of these stars ranges from more compact (${\sim}30~R_\odot$) for LBVs to more extended (${\sim}500~R_\odot$) for YHG and RSG progenitors, and their temperatures range from 3700--20000~K. The temperature and radii of existing IIb progenitors are roughly consistent with this range of parameters (see Figure \ref{fig:tracks}). Given the radius and temperature measured for \SN{}, and the observed variability, it is possible that \SN{} arose from an LBV progenitor. LBV stars are known to exhibit variability due to S-Dor type instability and more rarely can have Giant Eruptions \citep{Humphreys1994}. Galactic LBVs also span a wide range in luminosity and effective temperature (see Figure 13 in \citet{Groh2013}). According to \citet{Groh2013}, this spread is due to a range of initial masses with stars in the 20--25~$M_\odot$ range ending their lives as SNe and those with masses $>25~M_\odot$ continuing their blueward evolution to the Wolf-Rayet phase. LBVs were originally proposed as possible SN progenitors due to modulations observed in the radio light curves of SNe 2001ig and 2003bg, both of which were SNe IIb \citep{Kotak2006}, so there is observational support for LBVs as progenitors of IIb SNe. One line of evidence which may count against an LBV progenitor for \SN{} is that the observed luminosity of $\log(L/L_\odot)=4.84\pm0.14$ is below the range of luminosities observed for Galactic LBVs which roughly span 5.2--6.5 in $\log(L/L_\odot)$, although there are significant uncertainties on the luminosities due to uncertainties in their distances. The slightly higher luminosities for observed LBVs might explain the discrepancy between their initial masses of 20--25~$M_\odot$ compared to the lower mass range of 11--15$~M_\odot$ we obtained from single star models in Section \ref{subsec:singlestar}. 


Binary models are preferred over single star models due to the fact that single massive stars are less probable than binaries \citep{Sana2012}, while population studies indicate that a sufficient number of stripped-envelope progenitors are produced in close binary systems \citep{Podsiadlowski1993,Yoon2010,Eldridge2013}. Figure \ref{fig:BPASS} demonstrates that there are a large number of binary models whose endpoints in the H-R diagram sit within the yellow void occupied by \SN{}'s progenitor as well as those of other SNe IIb progenitors.  The strongest evidence for binary systems, however, comes from the direct detection of binary companions for SNe 1993J and 2011dh \citep{Maund2004,Fox2014,Folatelli2014,Maund2019}. Our preferred BPASS model consisted of a $12+1.2~M_\odot$ binary system which underwent unstable mass transfer during a common envelope phase in which most of the hydrogen envelope of the primary was lost from the system. Other high-probability BPASS systems consistent with \SN{}'s progenitor have primaries in the range 10--12~$M_\odot$ and secondaries in the range 1--3~$M_\odot$. These systems are quite different to the $15+14~M_\odot$ system that led to SN~1993J \citep{Maund2004} and the $16+10~M_\odot$ system responsible for SN~2011dh \citep{Benvenuto2013}. The reason is that the \SN{} binary systems have mass ratios $q\lesssim0.5$ which undergo an unstable mass transfer phase, whereas the systems considered by \citet{Claeys2011} and \citet{Benvenuto2013} only undergo stable mass transfer. The bulk of the mass-loss and stripping of the H-rich envelope in \SN{} therefore happened during the period of unstable mass transfer rather than during stable mass transfer. For the stable mass transfer systems considered by \citet{Claeys2011}, they found that 93\% of binaries are expected to have O or B-type companion stars, in agreement with the hot companions observed for SNe 1993J and 2011dh. A deep search for a possible surviving companion to \SN{} is therefore required to definitively determine the nature of the binary system, if it is indeed a binary, since the detection of a hot companion star would rule out our BPASS models. The non-detection of a companion will not rule out our $12+1.2~M_\odot$ BPASS model, however. This is because the companion is a roughly solar-type star with a luminosity of $\log(L/L_{\odot})=0.34$, which is approximately four orders of magnitude fainter than the progenitor candidate's. The corresponding apparent magnitude of $R{\approx}32$~mag would be too faint even for \HST{}. 

An intriguing hypothesis put forward by \citet{Maeda2015} is that the timing of a binary interaction determines the mass-loss rate and extent of the progenitor's hydrogen-rich envelope. SNe IIb with double-peaked light curves such as SNe 1993J and 2013df were found to show broad and flat-topped H$\alpha$ and \ion{He}{i} emission lines in their spectra more than 1 year post-explosion, which was taken as a clear sign of ejecta-circumstellar medium (CSM) interaction, in agreement with X-ray observations. They found a correlation between the CSM density and the early-time shock-cooling luminosity, with double-peaked SNe having higher CSM densities than those with a weak or absent shock-cooling peak. \citet{Maeda2015} also compared the derived mass-loss rates with the envelope radius and found that the double-peaked SNe have higher mass-loss rates and larger radii compared to single-peaked SNe. In our preferred BPASS model, the final phase of binary-driven mass loss begins $\approx4000$ years prior to core-collapse. This is early enough to agree with \citet{Maeda2015}'s hypothesis that a strong binary interaction phase well before the explosion ($\gtrsim 1000$~yr) can explain the weak or absent shock-cooling emission like we see for \SN{}. Furthermore, the actual mass-loss rate from the BPASS model of ${\sim}10^{-5}~M_\odot$/yr is low enough to be consistent with the delayed binary interaction group (see their Figure 6). The mass of the H-rich envelope decreases from ${\approx}0.02$ to ${\approx}0.01~M_\odot$ over this period. This reduction in the envelope mass could be crucial in preventing a prominent shock-cooling peak as demonstrated in the mass-radius parameter space in Figure \ref{fig:bol_model}. Looking forward, we recommend late-time observations to monitor the light curve of \SN{} in order to search for a flattening caused by CSM-interaction. We also recommend spectral observations to search for box-liek H$\alpha$ emission from a CSM-ejecta interaction. The absence of these two features would support the hypothesis that \SN{} underwent a binary interaction well before core-collapse, while the opposite case would force us to revise this interpretation. Analysis of the X-ray and radio data for \SN{} \citep{2024TNSAN.343,ATCA_atel} would also shed light on the CSM properties of the system.

\section{Conclusions}

In this paper we conducted a study of the progenitor candidate of the Type IIb \SN{}, the seventh SN IIb with a direct progenitor detection. We also studied the SN itself with the aim of connecting the progenitor star properties with the properties of the ensuing explosion. We summarise our conclusions here:
\begin{enumerate}

\item The position of \SN{} in our ERIS adaptive optics imaging agrees with the position of the progenitor candidate identified in archival \HST{} imaging within 19~mas. This makes it highly likely that the progenitor candidate is the actual progenitor to \SN{}, although late-time \HST{} imaging to search for the disappearance will be required to confirm the association.

\item Archival DECam imaging in the $griz$ bands allowed us to construct the SED of the progenitor candidate. The SED is consistent with a star having an A5 spectral type, a temperature of $8400^{+400}_{500}$~K, a radius of $124^{+18}_{-8}~R_\odot$ and a luminosity of $\mathrm{log}(L/L_{\odot})=4.84\pm0.14$. From single-star models we estimate an initial mass of $13\pm2~M_\odot$ for the progenitor candidate, while with BPASS models we find preference for a $12+1.2~M_\odot$ binary system with an initial period of 1.73 years. 

\item We find evidence for significant variability of the progenitor candidate in archival $gr$ DECam images, as well as from the \HST{}/WFPC2 F814W flux in 2001 compared to the DECam $i$-band brightness between 2013 and 2016, which showed a decline of 0.7~mag. 

\item \SN{} is the least-luminous Type IIb SN with a progenitor detection. The $r$-band light curve deviates from other Type IIb SNe with progenitor detections at $>60$ days with the rate of decline after 60 days being shallower (0.013~mag/day) than the comparison sample (0.022 mag/day) but still not as shallow as the 0.0098 mag/day expected from complete \textsuperscript{56}Co energy deposition. 
\item We modelled the bolometric light curve of \SN{} with the semi-analytic core and shell model of \citet{Nagy2016} and derived a synthesised \textsuperscript{56}Ni mass of $0.055{~}M_\odot$, an ejecta mass of $2.55~M_\odot$, and a kinetic energy of $1.95\times10^{51}$~erg, assuming an opacity of $\kappa=0.2$~cm$^2$~g$^{-1}$ in the core component. Our model requires $\gamma$-ray leakage in order to match the late-time luminosity, but the decay rate is not consistent with the observed decay rate. More detailed modelling of the late-time light curve is required, which we leave to a future work. 

\item Although the early ($<5$~days) $g$, $B$ and $U$ light curves show evidence for a weak or short-lived shock-cooling phase, we were unable to constrain the radius and mass of the hydrogen-rich envelope via analytical shock-cooling models. Instead we used the \citet{Nagy2016} core and shell model to explore the allowed radius and mass parameter space for the envelope and found that the radius of the progenitor is essentially unconstrained for envelope masses $\lesssim0.006~M_\odot$. The radius of ${\approx}125~R_\odot$ we measured from the progenitor candidate SED is therefore consistent with our SN observations provided the envelope mass is less than $0.0065~M_\odot$. If, on the other hand, the envelope mass is $\approx0.01~M_\odot$, we would require the progenitor to be much more compact, with $R\lesssim6~R_\odot$. 

\item The earliest spectrum of \SN{} is most similar to that of SN~2008ax since both lack a prominent shock cooling phase in their light curves, whereas the first spectra of SNe 1993J, 2013df and 2016gkg were all blue and featureless due to shock-cooling emission. The continued strength of H$\alpha$ distinguishes \SN{} from other Type IIb SNe with progenitor detections, and is most clearly apparent in the spectra at ${\sim}80$ days where the pEW of H$\alpha$ is twice as large or more compared to SNe 1993J, 2008ax and 2011dh at the same epoch. 

\item Although single-star models can explain the progenitor and subsequent supernova, we prefer a binary origin due to the statistical likelihood of massive stars being in binaries, the precedent of two previous direct detections of binary companions for SNe 1993J and 2011dh, and the larger number of binary models with endpoints terminating at the progenitor's position in the H-R diagram. 

\item We recommend late-time \HST{} imaging to confirm the disappearance of the progenitor candidate and to search for a possible binary companion. A hot binary companion would rule out our BPASS model and models which undergo a period of unstable mass transfer, forcing us to revise the analysis presented here. We also recommend spectroscopic observations at times $\gtrsim1$~yr to search for signs of CSM-ejecta interaction. The absence of such a signature would agree with \citet{Maeda2015}'s hypothesis that single-peaked SNe IIb result from binaries which underwent a period of mass transfer well before ($\gtrsim1000$~yr) tnhe explosion, although this would still not rule out single-star models. 

\end{enumerate}

\begin{acknowledgements}
GL and SdW were supported by a research grant (VIL60862) from VILLUM FONDEN. This work made use of observations made at the South African Astronomical Observatory (SAAO) that is financially supported by the South African National Research Foundation (NRF). Some of the observations reported in this paper were obtained with the Southern African Large Telescope (SALT), under programme 2024-2-LSP-001 (PI: DAHB). Polish participation in SALT is funded by grant No. MNiSW DIR/WK/2016/07. This research uses services or data provided by the SPectra Analysis and Retrievable Catalog Lab (SPARCL) and the Astro Data Lab, which are both part of the Community Science and Data Center (CSDC) Program of NSF NOIRLab. NOIRLab is operated by the Association of Universities for Research in Astronomy (AURA), Inc. under a cooperative agreement with the U.S. National Science Foundation. This work made use of data supplied by the UK Swift Science Data Centre at the University of Leicester. This research made use of Photutils, an Astropy package for detection and photometry of astronomical sources \citep{photutils}.
\end{acknowledgements}
\bibliographystyle{aa}
\bibliography{aas}

\begin{thebibliography}{101}
\expandafter\ifx\csname natexlab\endcsname\relax\def\natexlab#1{#1}\fi

\bibitem[{{Abbott} {et~al.}(2018){Abbott}, {Abdalla}, {Allam}, {Amara},
  {Annis}, {Asorey}, {Avila}, {Ballester}, {Banerji}, {Barkhouse}, {Baruah},
  {Baumer}, {Bechtol}, {Becker}, {Benoit-L{\'e}vy}, {Bernstein}, {Bertin},
  {Blazek}, {Bocquet}, {Brooks}, {Brout}, {Buckley-Geer}, {Burke}, {Busti},
  {Campisano}, {Cardiel-Sas}, {Carnero Rosell}, {Carrasco Kind}, {Carretero},
  {Castander}, {Cawthon}, {Chang}, {Chen}, {Conselice}, {Costa}, {Crocce},
  {Cunha}, {D'Andrea}, {da Costa}, {Das}, {Daues}, {Davis}, {Davis}, {De
  Vicente}, {DePoy}, {DeRose}, {Desai}, {Diehl}, {Dietrich}, {Dodelson},
  {Doel}, {Drlica-Wagner}, {Eifler}, {Elliott}, {Evrard}, {Farahi}, {Fausti
  Neto}, {Fernandez}, {Finley}, {Flaugher}, {Foley}, {Fosalba}, {Friedel},
  {Frieman}, {Garc{\'\i}a-Bellido}, {Gaztanaga}, {Gerdes}, {Giannantonio},
  {Gill}, {Glazebrook}, {Goldstein}, {Gower}, {Gruen}, {Gruendl}, {Gschwend},
  {Gupta}, {Gutierrez}, {Hamilton}, {Hartley}, {Hinton}, {Hislop}, {Hollowood},
  {Honscheid}, {Hoyle}, {Huterer}, {Jain}, {James}, {Jeltema}, {Johnson},
  {Johnson}, {Kacprzak}, {Kent}, {Khullar}, {Klein}, {Kovacs}, {Koziol},
  {Krause}, {Kremin}, {Kron}, {Kuehn}, {Kuhlmann}, {Kuropatkin}, {Lahav},
  {Lasker}, {Li}, {Li}, {Liddle}, {Lima}, {Lin}, {L{\'o}pez-Reyes}, {MacCrann},
  {Maia}, {Maloney}, {Manera}, {March}, {Marriner}, {Marshall}, {Martini},
  {McClintock}, {McKay}, {McMahon}, {Melchior}, {Menanteau}, {Miller},
  {Miquel}, {Mohr}, {Morganson}, {Mould}, {Neilsen}, {Nichol}, {Nogueira},
  {Nord}, {Nugent}, {Nunes}, {Ogando}, {Old}, {Pace}, {Palmese},
  {Paz-Chinch{\'o}n}, {Peiris}, {Percival}, {Petravick}, {Plazas}, {Poh},
  {Pond}, {Porredon}, {Pujol}, {Refregier}, {Reil}, {Ricker}, {Rollins},
  {Romer}, {Roodman}, {Rooney}, {Ross}, {Rykoff}, {Sako}, {Sanchez}, {Sanchez},
  {Santiago}, {Saro}, {Scarpine}, {Scolnic}, {Serrano}, {Sevilla-Noarbe},
  {Sheldon}, {Shipp}, {Silveira}, {Smith}, {Smith}, {Smith}, {Soares-Santos},
  {Sobreira}, {Song}, {Stebbins}, {Suchyta}, {Sullivan}, {Swanson}, {Tarle},
  {Thaler}, {Thomas}, {Thomas}, {Troxel}, {Tucker}, {Vikram}, {Vivas},
  {Walker}, {Wechsler}, {Weller}, {Wester}, {Wolf}, {Wu}, {Yanny}, {Zenteno},
  {Zhang}, {Zuntz}, {DES Collaboration}, {Juneau}, {Fitzpatrick}, \&
  {Nikutta}}]{DES2018}
{Abbott}, T.~M.~C., {Abdalla}, F.~B., {Allam}, S., {et~al.} 2018, \apjs, 239,
  18

\bibitem[{{Aldering} {et~al.}(1994){Aldering}, {Humphreys}, \&
  {Richmond}}]{Aldering1994}
{Aldering}, G., {Humphreys}, R.~M., \& {Richmond}, M. 1994, \aj, 107, 662

\bibitem[{{Arcavi} {et~al.}(2011){Arcavi}, {Gal-Yam}, {Yaron}, {Sternberg},
  {Rabinak}, {Waxman}, {Kasliwal}, {Quimby}, {Ofek}, {Horesh}, {Kulkarni},
  {Filippenko}, {Silverman}, {Cenko}, {Li}, {Bloom}, {Sullivan}, {Nugent},
  {Poznanski}, {Gorbikov}, {Fulton}, {Howell}, {Bersier}, {Riou},
  {Lamotte-Bailey}, {Griga}, {Cohen}, {Hachinger}, {Polishook}, {Xu},
  {Ben-Ami}, {Manulis}, {Walker}, {Maguire}, {Pan}, {Matheson}, {Mazzali},
  {Pian}, {Fox}, {Gehrels}, {Law}, {James}, {Marchant}, {Smith}, {Mottram},
  {Barnsley}, {Kandrashoff}, \& {Clubb}}]{Arcavi2011}
{Arcavi}, I., {Gal-Yam}, A., {Yaron}, O., {et~al.} 2011, \apjl, 742, L18

\bibitem[{{Arcavi} {et~al.}(2017){Arcavi}, {Hosseinzadeh}, {Brown}, {Smartt},
  {Valenti}, {Tartaglia}, {Piro}, {Sanchez}, {Nicholls}, {Monard}, {Howell},
  {McCully}, {Sand}, {Tonry}, {Denneau}, {Stalder}, {Heinze}, {Rest}, {Smith},
  \& {Bishop}}]{Arcavi2017}
{Arcavi}, I., {Hosseinzadeh}, G., {Brown}, P.~J., {et~al.} 2017, \apjl, 837, L2

\bibitem[{{Arnett} \& {Fu}(1989)}]{Arnett1989}
{Arnett}, W.~D. \& {Fu}, A. 1989, \apj, 340, 396

\bibitem[{{Benvenuto} {et~al.}(2013){Benvenuto}, {Bersten}, \&
  {Nomoto}}]{Benvenuto2013}
{Benvenuto}, O.~G., {Bersten}, M.~C., \& {Nomoto}, K. 2013, \apj, 762, 74

\bibitem[{{Bersten} {et~al.}(2012){Bersten}, {Benvenuto}, {Nomoto}, {Ergon},
  {Folatelli}, {Sollerman}, {Benetti}, {Botticella}, {Fraser}, {Kotak},
  {Maeda}, {Ochner}, \& {Tomasella}}]{Bersten2012}
{Bersten}, M.~C., {Benvenuto}, O.~G., {Nomoto}, K., {et~al.} 2012, \apj, 757,
  31

\bibitem[{{Blinnikov} {et~al.}(1998){Blinnikov}, {Eastman}, {Bartunov},
  {Popolitov}, \& {Woosley}}]{Blinnikov1998}
{Blinnikov}, S.~I., {Eastman}, R., {Bartunov}, O.~S., {Popolitov}, V.~A., \&
  {Woosley}, S.~E. 1998, \apj, 496, 454

\bibitem[{{B{\"o}ker} {et~al.}(2002){B{\"o}ker}, {Laine}, {van der Marel},
  {Sarzi}, {Rix}, {Ho}, \& {Shields}}]{Boeker2002}
{B{\"o}ker}, T., {Laine}, S., {van der Marel}, R.~P., {et~al.} 2002, \aj, 123,
  1389

\bibitem[{Bradley {et~al.}(2024)Bradley, Sip{\H o}cz, Robitaille, Tollerud,
  Vin{\'{\i}}cius, Deil, Barbary, Wilson, Busko, Donath, G{\"u}nther, Cara,
  Lim, Me{\ss}linger, Conseil, Burnett, Bostroem, Droettboom, Bray, Bratholm,
  Ginsburg, Jamieson, Barentsen, Craig, Morris, Perrin, Rathi, Pascual, \&
  Georgiev}]{photutils}
Bradley, L., Sip{\H o}cz, B., Robitaille, T., {et~al.} 2024, astropy/photutils:
  2.0.2

\bibitem[{{Buckley} {et~al.}(2006){Buckley}, {Swart}, \&
  {Meiring}}]{Buckley2006}
{Buckley}, D. A.~H., {Swart}, G.~P., \& {Meiring}, J.~G. 2006, in Society of
  Photo-Optical Instrumentation Engineers (SPIE) Conference Series, Vol. 6267,
  Ground-based and Airborne Telescopes, ed. L.~M. {Stepp}, 62670Z

\bibitem[{{Campana} {et~al.}(2024){Campana}, {Asquini}, {Smartt}, {Izzo},
  {Landoni}, {D'Avanzo}, {Young}, {Smith}, {Ben-Ami}, \&
  {Ofek}}]{2024TNSAN.343}
{Campana}, S., {Asquini}, L., {Smartt}, S., {et~al.} 2024, Transient Name
  Server AstroNote, 343, 1

\bibitem[{{Cardelli} {et~al.}(1989){Cardelli}, {Clayton}, \&
  {Mathis}}]{Cardelli1989}
{Cardelli}, J.~A., {Clayton}, G.~C., \& {Mathis}, J.~S. 1989, \apj, 345, 245

\bibitem[{{Chevalier} \& {Soderberg}(2010)}]{Chevalier2010}
{Chevalier}, R.~A. \& {Soderberg}, A.~M. 2010, \apjl, 711, L40

\bibitem[{{Choi} {et~al.}(2016){Choi}, {Dotter}, {Conroy}, {Cantiello},
  {Paxton}, \& {Johnson}}]{Choi2016}
{Choi}, J., {Dotter}, A., {Conroy}, C., {et~al.} 2016, \apj, 823, 102

\bibitem[{{Claeys} {et~al.}(2011){Claeys}, {de Mink}, {Pols}, {Eldridge}, \&
  {Baes}}]{Claeys2011}
{Claeys}, J.~S.~W., {de Mink}, S.~E., {Pols}, O.~R., {Eldridge}, J.~J., \&
  {Baes}, M. 2011, \aap, 528, A131

\bibitem[{{Crockett} {et~al.}(2008){Crockett}, {Eldridge}, {Smartt},
  {Pastorello}, {Gal-Yam}, {Fox}, {Leonard}, {Kasliwal}, {Mattila}, {Maund},
  {Stephens}, \& {Danziger}}]{Crockett2008}
{Crockett}, R.~M., {Eldridge}, J.~J., {Smartt}, S.~J., {et~al.} 2008, \mnras,
  391, L5

\bibitem[{{Davies} {et~al.}(2023){Davies}, {Absil}, {Agapito}, {Agudo Berbel},
  {Baruffolo}, {Biliotti}, {Black}, {Bonaglia}, {Bonse}, {Briguglio},
  {Campana}, {Cao}, {Carbonaro}, {Cortes}, {Cresci}, {Dallilar}, {Dannert}, {De
  Rosa}, {Deysenroth}, {Di Antonio}, {Di Cianno}, {Di Rico}, {Doelman},
  {Dolci}, {Dorn}, {Eisenhauer}, {Esposito}, {Fantinel}, {Ferruzzi},
  {Feuchtgruber}, {Finger}, {F{\"o}rster Schreiber}, {Gao}, {Gemperlein},
  {Genzel}, {Gillessen}, {Ginski}, {Glauser}, {Glindemann}, {Grani}, {Hartl},
  {Hayoz}, {Heida}, {Henry}, {Hofmann}, {Huber}, {Kasper}, {Keller},
  {Kenworthy}, {Kravchenko}, {Kuntschner}, {Lacour}, {Lightfoot}, {Lunney},
  {Lutz}, {Macintosh}, {Mannucci}, {Marsset}, {Modigliani}, {Neeser}, {Orban de
  Xivry}, {Ott}, {Pallanca}, {Patapis}, {Pearson}, {Pe{\~n}a}, {Percheron},
  {Puglisi}, {Quanz}, {Rabien}, {Rau}, {Riccardi}, {Salasnich}, {Schmid},
  {Schubert}, {Serra}, {Shimizu}, {Snik}, {Sturm}, {Tacconi}, {Taylor},
  {Valentini}, {Waring}, {Wiezorrek}, \& {Xompero}}]{Davies2023}
{Davies}, R., {Absil}, O., {Agapito}, G., {et~al.} 2023, \aap, 674, A207

\bibitem[{{de Wet} {et~al.}(2024){de Wet}, {Zimmerman}, \&
  {Erasmus}}]{2024TNSAN.342}
{de Wet}, S., {Zimmerman}, E., \& {Erasmus}, N. 2024, Transient Name Server
  AstroNote, 342, 1

\bibitem[{{Dessart} {et~al.}(2011){Dessart}, {Hillier}, {Livne}, {Yoon},
  {Woosley}, {Waldman}, \& {Langer}}]{Dessart2011}
{Dessart}, L., {Hillier}, D.~J., {Livne}, E., {et~al.} 2011, \mnras, 414, 2985

\bibitem[{{Dessart} {et~al.}(2018){Dessart}, {Yoon}, {Livne}, \&
  {Waldman}}]{Dessart2018}
{Dessart}, L., {Yoon}, S.-C., {Livne}, E., \& {Waldman}, R. 2018, \aap, 612,
  A61

\bibitem[{{Dolphin}(2016)}]{Dolphin2016}
{Dolphin}, A. 2016, {DOLPHOT: Stellar photometry}, Astrophysics Source Code
  Library, record ascl:1608.013

\bibitem[{{Dolphin}(2000)}]{Dolphin2000}
{Dolphin}, A.~E. 2000, \pasp, 112, 1383

\bibitem[{{Dotter}(2016)}]{Dotter2016}
{Dotter}, A. 2016, \apjs, 222, 8

\bibitem[{{Eldridge} {et~al.}(2013){Eldridge}, {Fraser}, {Smartt}, {Maund}, \&
  {Crockett}}]{Eldridge2013}
{Eldridge}, J.~J., {Fraser}, M., {Smartt}, S.~J., {Maund}, J.~R., \&
  {Crockett}, R.~M. 2013, \mnras, 436, 774

\bibitem[{{Eldridge} {et~al.}(2017){Eldridge}, {Stanway}, {Xiao}, {McClelland},
  {Taylor}, {Ng}, {Greis}, \& {Bray}}]{Eldridge2017}
{Eldridge}, J.~J., {Stanway}, E.~R., {Xiao}, L., {et~al.} 2017, \pasa, 34, e058

\bibitem[{{Erasmus} {et~al.}(2024){Erasmus}, {Steele}, {Piascik}, {Bates},
  {Mottram}, {Rosie}, {van Gend}, {Geen}, {Pretorius}, {Potter}, {Loubser},
  {Koorts}, {Gajjar}, {Titus}, {Worters}, {Sickafoose}, {Chandra}, {O'Connor},
  {Matlala}, {Crook-Mansour}, {Ranjbar}, {Smith}, {Jermak}, {Abiodun}, \&
  {Egbo}}]{Erasmus2024}
{Erasmus}, N., {Steele}, I.~A., {Piascik}, A.~S., {et~al.} 2024, Journal of
  Astronomical Telescopes, Instruments, and Systems, 10, 025005

\bibitem[{{Ergon} {et~al.}(2015){Ergon}, {Jerkstrand}, {Sollerman},
  {Elias-Rosa}, {Fransson}, {Fraser}, {Pastorello}, {Kotak}, {Taubenberger},
  {Tomasella}, {Valenti}, {Benetti}, {Helou}, {Kasliwal}, {Maund}, {Smartt}, \&
  {Spyromilio}}]{Ergon2015}
{Ergon}, M., {Jerkstrand}, A., {Sollerman}, J., {et~al.} 2015, \aap, 580, A142

\bibitem[{{Filippenko}(1997)}]{Filippenko1997}
{Filippenko}, A.~V. 1997, \araa, 35, 309

\bibitem[{{Fitzpatrick} {et~al.}(2014){Fitzpatrick}, {Olsen}, {Economou},
  {Stobie}, {Beers}, {Dickinson}, {Norris}, {Saha}, {Seaman}, {Silva},
  {Swaters}, {Thomas}, \& {Valdes}}]{Fitzpatrick2014}
{Fitzpatrick}, M.~J., {Olsen}, K., {Economou}, F., {et~al.} 2014, in Society of
  Photo-Optical Instrumentation Engineers (SPIE) Conference Series, Vol. 9149,
  Observatory Operations: Strategies, Processes, and Systems V, ed. A.~B.
  {Peck}, C.~R. {Benn}, \& R.~L. {Seaman}, 91491T

\bibitem[{{Flaugher} {et~al.}(2015){Flaugher}, {Diehl}, {Honscheid}, {Abbott},
  {Alvarez}, {Angstadt}, {Annis}, {Antonik}, {Ballester}, {Beaufore},
  {Bernstein}, {Bernstein}, {Bigelow}, {Bonati}, {Boprie}, {Brooks},
  {Buckley-Geer}, {Campa}, {Cardiel-Sas}, {Castander}, {Castilla}, {Cease},
  {Cela-Ruiz}, {Chappa}, {Chi}, {Cooper}, {da Costa}, {Dede}, {Derylo},
  {DePoy}, {de Vicente}, {Doel}, {Drlica-Wagner}, {Eiting}, {Elliott}, {Emes},
  {Estrada}, {Fausti Neto}, {Finley}, {Flores}, {Frieman}, {Gerdes},
  {Gladders}, {Gregory}, {Gutierrez}, {Hao}, {Holland}, {Holm}, {Huffman},
  {Jackson}, {James}, {Jonas}, {Karcher}, {Karliner}, {Kent}, {Kessler},
  {Kozlovsky}, {Kron}, {Kubik}, {Kuehn}, {Kuhlmann}, {Kuk}, {Lahav}, {Lathrop},
  {Lee}, {Levi}, {Lewis}, {Li}, {Mandrichenko}, {Marshall}, {Martinez},
  {Merritt}, {Miquel}, {Mu{\~n}oz}, {Neilsen}, {Nichol}, {Nord}, {Ogando},
  {Olsen}, {Palaio}, {Patton}, {Peoples}, {Plazas}, {Rauch}, {Reil}, {Rheault},
  {Roe}, {Rogers}, {Roodman}, {Sanchez}, {Scarpine}, {Schindler}, {Schmidt},
  {Schmitt}, {Schubnell}, {Schultz}, {Schurter}, {Scott}, {Serrano}, {Shaw},
  {Smith}, {Soares-Santos}, {Stefanik}, {Stuermer}, {Suchyta}, {Sypniewski},
  {Tarle}, {Thaler}, {Tighe}, {Tran}, {Tucker}, {Walker}, {Wang}, {Watson},
  {Weaverdyck}, {Wester}, {Woods}, {Yanny}, \& {DES Collaboration}}]{DECAM}
{Flaugher}, B., {Diehl}, H.~T., {Honscheid}, K., {et~al.} 2015, \aj, 150, 150

\bibitem[{{Folatelli} {et~al.}(2014){Folatelli}, {Bersten}, {Benvenuto}, {Van
  Dyk}, {Kuncarayakti}, {Maeda}, {Nozawa}, {Nomoto}, {Hamuy}, \&
  {Quimby}}]{Folatelli2014}
{Folatelli}, G., {Bersten}, M.~C., {Benvenuto}, O.~G., {et~al.} 2014, \apjl,
  793, L22

\bibitem[{{Folatelli} {et~al.}(2015){Folatelli}, {Bersten}, {Kuncarayakti},
  {Benvenuto}, {Maeda}, \& {Nomoto}}]{Folatelli2015}
{Folatelli}, G., {Bersten}, M.~C., {Kuncarayakti}, H., {et~al.} 2015, \apj,
  811, 147

\bibitem[{{Fox} {et~al.}(2014){Fox}, {Azalee Bostroem}, {Van Dyk},
  {Filippenko}, {Fransson}, {Matheson}, {Cenko}, {Chandra}, {Dwarkadas}, {Li},
  {Parker}, \& {Smith}}]{Fox2014}
{Fox}, O.~D., {Azalee Bostroem}, K., {Van Dyk}, S.~D., {et~al.} 2014, \apj,
  790, 17

\bibitem[{{Groh} {et~al.}(2013){Groh}, {Meynet}, {Georgy}, \&
  {Ekstr{\"o}m}}]{Groh2013}
{Groh}, J.~H., {Meynet}, G., {Georgy}, C., \& {Ekstr{\"o}m}, S. 2013, \aap,
  558, A131

\bibitem[{{Groot} {et~al.}(2024){Groot}, {Bloemen}, {Vreeswijk}, {van Roestel},
  {Jonker}, {Nelemans}, {Klein-Wolt}, {Lepoole}, {Pieterse}, {Rodenhuis},
  {Boland}, {Haverkorn}, {Aerts}, {Bakker}, {Balster}, {Bekema}, {Dijkstra},
  {Dolron}, {Elswijk}, {van Elteren}, {Engels}, {Fokker}, {de Haan}, {Hahn},
  {ter Horst}, {Lesman}, {Kragt}, {Morren}, {Nillissen}, {Pessemier}, {Raskin},
  {de Rijke}, {Scheers}, {Schuil}, {Timmer}, {Antunes Amaral},
  {Arancibia-Rojas}, {Arcavi}, {Blagorodnova}, {Biswas}, {Breton}, {Dawson},
  {Dayal}, {De Wet}, {Duffy}, {Faris}, {Fausnaugh}, {Gal-Yam}, {Geier},
  {Horesh}, {Johnston}, {Katusiime}, {Kelley}, {Kosakowski}, {Kupfer},
  {Leloudas}, {Levan}, {Modiano}, {Mogawana}, {Munday}, {Paice}, {Patat},
  {Pelisoli}, {Ramsay}, {Ranaivomanana}, {Ruiz-Carmona}, {Schaffenroth},
  {Scaringi}, {Stoppa}, {Street}, {Tranin}, {Uzundag}, {Valenti},
  {Veresvarska}, {Vuc̆kovi{\'c}}, {Wichern}, {Wijers}, {Wijnands}, \&
  {Zimmerman}}]{Groot2024}
{Groot}, P.~J., {Bloemen}, S., {Vreeswijk}, P.~M., {et~al.} 2024, \pasp, 136,
  115003

\bibitem[{{Guti{\'e}rrez} {et~al.}(2017){Guti{\'e}rrez}, {Anderson}, {Hamuy},
  {Morrell}, {Gonz{\'a}lez-Gaitan}, {Stritzinger}, {Phillips}, {Galbany},
  {Folatelli}, {Dessart}, {Contreras}, {Della Valle}, {Freedman}, {Hsiao},
  {Krisciunas}, {Madore}, {Maza}, {Suntzeff}, {Prieto}, {Gonz{\'a}lez},
  {Cappellaro}, {Navarrete}, {Pizzella}, {Ruiz}, {Smith}, \&
  {Turatto}}]{Gutierrez2017}
{Guti{\'e}rrez}, C.~P., {Anderson}, J.~P., {Hamuy}, M., {et~al.} 2017, \apj,
  850, 89

\bibitem[{{Humphreys} \& {Davidson}(1994)}]{Humphreys1994}
{Humphreys}, R.~M. \& {Davidson}, K. 1994, \pasp, 106, 1025

\bibitem[{{Kilpatrick} {et~al.}(2022){Kilpatrick}, {Coulter}, {Foley}, {Piro},
  {Rest}, {Rojas-Bravo}, \& {Siebert}}]{Kilpatrick2022}
{Kilpatrick}, C.~D., {Coulter}, D.~A., {Foley}, R.~J., {et~al.} 2022, \apj,
  936, 111

\bibitem[{{Kilpatrick} {et~al.}(2017){Kilpatrick}, {Foley}, {Abramson}, {Pan},
  {Lu}, {Williams}, {Treu}, {Siebert}, {Fassnacht}, \& {Max}}]{Kilpatrick2017}
{Kilpatrick}, C.~D., {Foley}, R.~J., {Abramson}, L.~E., {et~al.} 2017, \mnras,
  465, 4650

\bibitem[{{Kilpatrick} {et~al.}(2018){Kilpatrick}, {Takaro}, {Foley},
  {Leibler}, {Pan}, {Campbell}, {Jacobson-Galan}, {Lewis}, {Lyke}, {Max},
  {Medallon}, \& {Rest}}]{Kilpatrick2018}
{Kilpatrick}, C.~D., {Takaro}, T., {Foley}, R.~J., {et~al.} 2018, \mnras, 480,
  2072

\bibitem[{{Kotak} \& {Vink}(2006)}]{Kotak2006}
{Kotak}, R. \& {Vink}, J.~S. 2006, \aap, 460, L5

\bibitem[{{Kumar} {et~al.}(2013){Kumar}, {Pandey}, {Sahu}, {Vinko},
  {Moskvitin}, {Anupama}, {Bhatt}, {Ordasi}, {Nagy}, {Sokolov}, {Sokolova},
  {Komarova}, {Kumar}, {Bose}, {Roy}, \& {Sagar}}]{Kumar2013}
{Kumar}, B., {Pandey}, S.~B., {Sahu}, D.~K., {et~al.} 2013, \mnras, 431, 308

\bibitem[{{Lam} {et~al.}(2023){Lam}, {Smith}, {Arcavi}, {Steele},
  {Veitch-Michaelis}, \& {Wyrzykowski}}]{Lam2023}
{Lam}, M.~C., {Smith}, R.~J., {Arcavi}, I., {et~al.} 2023, \aj, 166, 13

\bibitem[{{Lewis} {et~al.}(1994){Lewis}, {Walton}, {Meikle}, {Martin},
  {Cumming}, {Catchpole}, {Arevalo}, {Argyle}, {Benn}, {Bunclark}, {Castaneda},
  {Centurion}, {Clegg}, {Delgado}, {Dhillon}, {Goudfrooij}, {Harlaftis},
  {Hassall}, {Helmer}, {Hill}, {Jones}, {King}, {Lazaro}, {Lucey}, {Martin},
  {Miller}, {Morrison}, {Penny}, {Perez}, {Read}, {Rudd}, {Rutten}, {Sharples},
  {Unger}, \& {Vilchez}}]{Lewis1994}
{Lewis}, J.~R., {Walton}, N.~A., {Meikle}, W.~P.~S., {et~al.} 1994, \mnras,
  266, L27

\bibitem[{{Liu} {et~al.}(2016){Liu}, {Modjaz}, {Bianco}, \& {Graur}}]{Liu2016}
{Liu}, Y.-Q., {Modjaz}, M., {Bianco}, F.~B., \& {Graur}, O. 2016, \apj, 827, 90

\bibitem[{{Lyman} {et~al.}(2014){Lyman}, {Bersier}, \& {James}}]{Lyman2014}
{Lyman}, J.~D., {Bersier}, D., \& {James}, P.~A. 2014, \mnras, 437, 3848

\bibitem[{{Maeda} {et~al.}(2015){Maeda}, {Hattori}, {Milisavljevic},
  {Folatelli}, {Drout}, {Kuncarayakti}, {Margutti}, {Kamble}, {Soderberg},
  {Tanaka}, {Kawabata}, {Kawabata}, {Yamanaka}, {Nomoto}, {Kim}, {Simon},
  {Phillips}, {Parrent}, {Nakaoka}, {Moriya}, {Suzuki}, {Takaki}, {Ishigaki},
  {Sakon}, {Tajitsu}, \& {Iye}}]{Maeda2015}
{Maeda}, K., {Hattori}, T., {Milisavljevic}, D., {et~al.} 2015, \apj, 807, 35

\bibitem[{{Maund}(2019)}]{Maund2019}
{Maund}, J.~R. 2019, \apj, 883, 86

\bibitem[{{Maund} {et~al.}(2011){Maund}, {Fraser}, {Ergon}, {Pastorello},
  {Smartt}, {Sollerman}, {Benetti}, {Botticella}, {Bufano}, {Danziger},
  {Kotak}, {Magill}, {Stephens}, \& {Valenti}}]{Maund2011dh}
{Maund}, J.~R., {Fraser}, M., {Ergon}, M., {et~al.} 2011, \apjl, 739, L37

\bibitem[{{Maund} {et~al.}(2004){Maund}, {Smartt}, {Kudritzki},
  {Podsiadlowski}, \& {Gilmore}}]{Maund2004}
{Maund}, J.~R., {Smartt}, S.~J., {Kudritzki}, R.~P., {Podsiadlowski}, P., \&
  {Gilmore}, G.~F. 2004, \nat, 427, 129

\bibitem[{{Milisavljevic} {et~al.}(2013){Milisavljevic}, {Margutti},
  {Soderberg}, {Pignata}, {Chomiuk}, {Fesen}, {Bufano}, {Sanders}, {Parrent},
  {Parker}, {Mazzali}, {Pian}, {Pickering}, {Buckley}, {Crawford}, {Gulbis},
  {Hettlage}, {Hooper}, {Nordsieck}, {O'Donoghue}, {Husser}, {Potter},
  {Kniazev}, {Kotze}, {Romero-Colmenero}, {Vaisanen}, {Wolf}, {Bietenholz},
  {Bartel}, {Fransson}, {Walker}, {Brunthaler}, {Chakraborti}, {Levesque},
  {MacFadyen}, {Drescher}, {Bock}, {Marples}, {Anderson}, {Benetti},
  {Reichart}, \& {Ivarsen}}]{Milisavljevic2013}
{Milisavljevic}, D., {Margutti}, R., {Soderberg}, A.~M., {et~al.} 2013, \apj,
  767, 71

\bibitem[{{Morag} {et~al.}(2023){Morag}, {Sapir}, \& {Waxman}}]{Morag2023}
{Morag}, J., {Sapir}, N., \& {Waxman}, E. 2023, \mnras, 522, 2764

\bibitem[{{Morales-Garoffolo} {et~al.}(2014){Morales-Garoffolo}, {Elias-Rosa},
  {Benetti}, {Taubenberger}, {Cappellaro}, {Pastorello}, {Klauser}, {Valenti},
  {Howerton}, {Ochner}, {Schramm}, {Siviero}, {Tartaglia}, \&
  {Tomasella}}]{Morales2014}
{Morales-Garoffolo}, A., {Elias-Rosa}, N., {Benetti}, S., {et~al.} 2014,
  \mnras, 445, 1647

\bibitem[{{Morozova} {et~al.}(2015){Morozova}, {Piro}, {Renzo}, {Ott},
  {Clausen}, {Couch}, {Ellis}, \& {Roberts}}]{Morozova2015}
{Morozova}, V., {Piro}, A.~L., {Renzo}, M., {et~al.} 2015, \apj, 814, 63

\bibitem[{{Nagy} {et~al.}(2014){Nagy}, {Ordasi}, {Vink{\'o}}, \&
  {Wheeler}}]{Nagy2014}
{Nagy}, A.~P., {Ordasi}, A., {Vink{\'o}}, J., \& {Wheeler}, J.~C. 2014, \aap,
  571, A77

\bibitem[{{Nagy} \& {Vink{\'o}}(2016)}]{Nagy2016}
{Nagy}, A.~P. \& {Vink{\'o}}, J. 2016, \aap, 589, A53

\bibitem[{{Nakar} \& {Piro}(2014)}]{Nakar2014}
{Nakar}, E. \& {Piro}, A.~L. 2014, \apj, 788, 193

\bibitem[{{Nakar} \& {Sari}(2010)}]{Nakar2010}
{Nakar}, E. \& {Sari}, R. 2010, \apj, 725, 904

\bibitem[{{Nikutta} {et~al.}(2020){Nikutta}, {Fitzpatrick}, {Scott}, \&
  {Weaver}}]{Nikutta2020}
{Nikutta}, R., {Fitzpatrick}, M., {Scott}, A., \& {Weaver}, B.~A. 2020,
  Astronomy and Computing, 33, 100411

\bibitem[{{Niu} {et~al.}(2024){Niu}, {Sun}, \& {Liu}}]{Niu2024}
{Niu}, Z., {Sun}, N.-C., \& {Liu}, J. 2024, \apjl, 970, L9

\bibitem[{{Niu} {et~al.}(2025){Niu}, {Sun}, {Maund}, {Guo}, {Li}, {Sun}, \&
  {Liu}}]{Niu2025}
{Niu}, Z., {Sun}, N.-C., {Maund}, J.~R., {et~al.} 2025, arXiv e-prints,
  arXiv:2504.20407

\bibitem[{{Pastorello} {et~al.}(2008){Pastorello}, {Kasliwal}, {Crockett},
  {Valenti}, {Arbour}, {Itagaki}, {Kaspi}, {Gal-Yam}, {Smartt}, {Griffith},
  {Maguire}, {Ofek}, {Seymour}, {Stern}, \& {Wiethoff}}]{Pastorello2008}
{Pastorello}, A., {Kasliwal}, M.~M., {Crockett}, R.~M., {et~al.} 2008, \mnras,
  389, 955

\bibitem[{{Pickles}(1998)}]{Pickles1998}
{Pickles}, A.~J. 1998, \pasp, 110, 863

\bibitem[{{Piro}(2015)}]{Piro2015}
{Piro}, A.~L. 2015, \apjl, 808, L51

\bibitem[{{Podsiadlowski} {et~al.}(1993){Podsiadlowski}, {Hsu}, {Joss}, \&
  {Ross}}]{Podsiadlowski1993}
{Podsiadlowski}, P., {Hsu}, J.~J.~L., {Joss}, P.~C., \& {Ross}, R.~R. 1993,
  \nat, 364, 509

\bibitem[{{Rabinak} \& {Waxman}(2011)}]{Rabinak2011}
{Rabinak}, I. \& {Waxman}, E. 2011, \apj, 728, 63

\bibitem[{{Reguitti} {et~al.}(2025){Reguitti}, {Pastorello}, {Smartt},
  {Valerin}, {Pignata}, {Campana}, {Chen}, {Sankar. K.}, {Moran}, {Mazzali},
  {Duarte}, {Salmaso}, {Anderson}, {Ashall}, {Benetti}, {Gromadzki},
  {Gutierrez}, {Humina}, {Inserra}, {Kankare}, {Kravtsov}, {Muller-Bravo},
  {Pessi}, {Sollerman}, {Young}, {Chambers}, {de Boer}, {Gao}, {Huber}, {Lin},
  {Lowe}, {Magnier}, {Minguez}, {Smith}, {Smith}, {Srivastav}, {Wainscoat}, \&
  {Benedet}}]{Reguitti2025}
{Reguitti}, A., {Pastorello}, A., {Smartt}, S.~J., {et~al.} 2025, arXiv
  e-prints, arXiv:2503.03851

\bibitem[{{Richmond} {et~al.}(1996){Richmond}, {Treffers}, {Filippenko}, \&
  {Paik}}]{Richmond1996}
{Richmond}, M.~W., {Treffers}, R.~R., {Filippenko}, A.~V., \& {Paik}, Y. 1996,
  \aj, 112, 732

\bibitem[{{Rolleston} {et~al.}(2000){Rolleston}, {Smartt}, {Dufton}, \&
  {Ryans}}]{Rolleston2000}
{Rolleston}, W.~R.~J., {Smartt}, S.~J., {Dufton}, P.~L., \& {Ryans}, R.~S.~I.
  2000, \aap, 363, 537

\bibitem[{{Roming} {et~al.}(2005){Roming}, {Kennedy}, {Mason}, {Nousek}, {Ahr},
  {Bingham}, {Broos}, {Carter}, {Hancock}, {Huckle}, {Hunsberger}, {Kawakami},
  {Killough}, {Koch}, {McLelland}, {Smith}, {Smith}, {Soto}, {Boyd},
  {Breeveld}, {Holland}, {Ivanushkina}, {Pryzby}, {Still}, \&
  {Stock}}]{Roming2005}
{Roming}, P. W.~A., {Kennedy}, T.~E., {Mason}, K.~O., {et~al.} 2005, \ssr, 120,
  95

\bibitem[{{Rose} {et~al.}(2024){Rose}, {Ryder}, {Maeda}, \&
  {Chandra}}]{ATCA_atel}
{Rose}, K., {Ryder}, S., {Maeda}, K., \& {Chandra}, P. 2024, The Astronomer's
  Telegram, 16920, 1

\bibitem[{{Rossa} {et~al.}(2006){Rossa}, {van der Marel}, {B{\"o}ker},
  {Gerssen}, {Ho}, {Rix}, {Shields}, \& {Walcher}}]{Rossa2006}
{Rossa}, J., {van der Marel}, R.~P., {B{\"o}ker}, T., {et~al.} 2006, \aj, 132,
  1074

\bibitem[{{Ryder} {et~al.}(2018){Ryder}, {Van Dyk}, {Fox}, {Zapartas}, {de
  Mink}, {Smith}, {Brunsden}, {Azalee Bostroem}, {Filippenko}, {Shivvers}, \&
  {Zheng}}]{Ryder2018}
{Ryder}, S.~D., {Van Dyk}, S.~D., {Fox}, O.~D., {et~al.} 2018, \apj, 856, 83

\bibitem[{{Sana} {et~al.}(2012){Sana}, {de Mink}, {de Koter}, {Langer},
  {Evans}, {Gieles}, {Gosset}, {Izzard}, {Le Bouquin}, \&
  {Schneider}}]{Sana2012}
{Sana}, H., {de Mink}, S.~E., {de Koter}, A., {et~al.} 2012, Science, 337, 444

\bibitem[{{Sapir} \& {Waxman}(2017)}]{Sapir2017}
{Sapir}, N. \& {Waxman}, E. 2017, \apj, 838, 130

\bibitem[{{Schlafly} \& {Finkbeiner}(2011)}]{SF2011}
{Schlafly}, E.~F. \& {Finkbeiner}, D.~P. 2011, \apj, 737, 103

\bibitem[{{Smartt}(2009)}]{Smartt2009}
{Smartt}, S.~J. 2009, \araa, 47, 63

\bibitem[{{Smartt}(2015)}]{Smartt2015}
{Smartt}, S.~J. 2015, \pasa, 32, e016

\bibitem[{{Smith} {et~al.}(2020){Smith}, {Smartt}, {Young}, {Tonry}, {Denneau},
  {Flewelling}, {Heinze}, {Weiland}, {Stalder}, {Rest}, {Stubbs}, {Anderson},
  {Chen}, {Clark}, {Do}, {F{\"o}rster}, {Fulton}, {Gillanders}, {McBrien},
  {O'Neill}, {Srivastav}, \& {Wright}}]{Smith2020}
{Smith}, K.~W., {Smartt}, S.~J., {Young}, D.~R., {et~al.} 2020, \pasp, 132,
  085002

\bibitem[{{Smith} {et~al.}(2024){Smith}, {Young}, {Nicholl}, {Fulton},
  {McCollum}, {Moore}, {Weston}, {Sheng}, {Aamer}, {Angus}, {Magill},
  {Ramsden}, {Shingles}, {Smartt}, {Srivastav}, {Gillanders}, {Stevance},
  {Cooper}, {Stoppa}, {Rhodes}, {Denneau}, {Tonry}, {Weiland}, {Siverd},
  {Erasmus}, {Koorts}, {Jordan}, {Suc}, {Rest}, {Chen}, {Stubbs}, {Sommer}, \&
  {Schmidt}}]{2024TNSAN.341}
{Smith}, K.~W., {Young}, D.~R., {Nicholl}, M., {et~al.} 2024, Transient Name
  Server AstroNote, 341, 1

\bibitem[{{Soderberg} {et~al.}(2012){Soderberg}, {Margutti}, {Zauderer},
  {Krauss}, {Katz}, {Chomiuk}, {Dittmann}, {Nakar}, {Sakamoto}, {Kawai},
  {Hurley}, {Barthelmy}, {Toizumi}, {Morii}, {Chevalier}, {Gurwell},
  {Petitpas}, {Rupen}, {Alexander}, {Levesque}, {Fransson}, {Brunthaler},
  {Bietenholz}, {Chugai}, {Grindlay}, {Copete}, {Connaughton}, {Briggs},
  {Meegan}, {von Kienlin}, {Zhang}, {Rau}, {Golenetskii}, {Mazets}, \&
  {Cline}}]{Soderberg2012}
{Soderberg}, A.~M., {Margutti}, R., {Zauderer}, B.~A., {et~al.} 2012, \apj,
  752, 78

\bibitem[{{Stanway} \& {Eldridge}(2018)}]{Stanway2018}
{Stanway}, E.~R. \& {Eldridge}, J.~J. 2018, \mnras, 479, 75

\bibitem[{{Stevance} {et~al.}(2020){Stevance}, {Eldridge}, \&
  {Stanway}}]{Stevance2020}
{Stevance}, H., {Eldridge}, J., \& {Stanway}, E. 2020, The Journal of Open
  Source Software, 5, 1987

\bibitem[{{Stevance} \& {Eldridge}(2021)}]{Stevance2021}
{Stevance}, H.~F. \& {Eldridge}, J.~J. 2021, \mnras, 504, L51

\bibitem[{{STScI Development Team}(2013)}]{pysynphot}
{STScI Development Team}. 2013, {pysynphot: Synthetic photometry software
  package}, Astrophysics Source Code Library, record ascl:1303.023

\bibitem[{{Szalai} {et~al.}(2016){Szalai}, {Vink{\'o}}, {Nagy}, {Silverman},
  {Wheeler}, {Dhungana}, {Marion}, {Kehoe}, {Fox}, {S{\'a}rneczky},
  {Marschalk{\'o}}, {B{\'\i}r{\'o}}, {Borkovits}, {Heged{\"u}s}, {Szak{\'a}ts},
  {Ferrante}, {B{\'a}nyai}, {Hodos{\'a}n}, {Kelemen}, \&
  {P{\'a}l}}]{Szalai2016}
{Szalai}, T., {Vink{\'o}}, J., {Nagy}, A.~P., {et~al.} 2016, \mnras, 460, 1500

\bibitem[{{Tartaglia} {et~al.}(2017){Tartaglia}, {Fraser}, {Sand}, {Valenti},
  {Smartt}, {McCully}, {Anderson}, {Arcavi}, {Elias-Rosa}, {Galbany},
  {Gal-Yam}, {Haislip}, {Hosseinzadeh}, {Howell}, {Inserra}, {Jha}, {Kankare},
  {Lundqvist}, {Maguire}, {Mattila}, {Reichart}, {Smith}, {Smith},
  {Stritzinger}, {Sullivan}, {Taddia}, \& {Tomasella}}]{Tartaglia2017}
{Tartaglia}, L., {Fraser}, M., {Sand}, D.~J., {et~al.} 2017, \apjl, 836, L12

\bibitem[{{Tonry} {et~al.}(2018){Tonry}, {Denneau}, {Heinze}, {Stalder},
  {Smith}, {Smartt}, {Stubbs}, {Weiland}, \& {Rest}}]{Tonry2018}
{Tonry}, J.~L., {Denneau}, L., {Heinze}, A.~N., {et~al.} 2018, \pasp, 130,
  064505

\bibitem[{{Tsvetkov} {et~al.}(2009){Tsvetkov}, {Volkov}, {Baklanov},
  {Blinnikov}, \& {Tuchin}}]{Tsevtkov2009}
{Tsvetkov}, D.~Y., {Volkov}, I.~M., {Baklanov}, P., {Blinnikov}, S., \&
  {Tuchin}, O. 2009, Peremennye Zvezdy, 29, 2

\bibitem[{{Van Dyk}(2017)}]{Vandyk2017}
{Van Dyk}, S.~D. 2017, Philosophical Transactions of the Royal Society of
  London Series A, 375, 20160277

\bibitem[{{Van Dyk} {et~al.}(2011){Van Dyk}, {Li}, {Cenko}, {Kasliwal},
  {Horesh}, {Ofek}, {Kraus}, {Silverman}, {Arcavi}, {Filippenko}, {Gal-Yam},
  {Quimby}, {Kulkarni}, {Yaron}, \& {Polishook}}]{Vandyk2011dh}
{Van Dyk}, S.~D., {Li}, W., {Cenko}, S.~B., {et~al.} 2011, \apjl, 741, L28

\bibitem[{{Van Dyk} {et~al.}(2018){Van Dyk}, {Zheng}, {Brink}, {Filippenko},
  {Milisavljevic}, {Andrews}, {Smith}, {Cignoni}, {Fox}, {Kelly}, {Adamo},
  {Yunus}, {Zhang}, \& {Kumar}}]{Vandyk2018}
{Van Dyk}, S.~D., {Zheng}, W., {Brink}, T.~G., {et~al.} 2018, \apj, 860, 90

\bibitem[{{Van Dyk} {et~al.}(2014){Van Dyk}, {Zheng}, {Fox}, {Cenko}, {Clubb},
  {Filippenko}, {Foley}, {Miller}, {Smith}, {Kelly}, {Lee}, {Ben-Ami}, \&
  {Gal-Yam}}]{Vandyk2014}
{Van Dyk}, S.~D., {Zheng}, W., {Fox}, O.~D., {et~al.} 2014, \aj, 147, 37

\bibitem[{{Van Dyk} {et~al.}(2019){Van Dyk}, {Zheng}, {Maund}, {Brink},
  {Srinivasan}, {Andrews}, {Smith}, {Leonard}, {Morozova}, {Filippenko},
  {Conner}, {Milisavljevic}, {de Jaeger}, {Long}, {Isaacson}, {Crossfield},
  {Kosiarek}, {Howard}, {Fox}, {Kelly}, {Piro}, {Littlefair}, {Dhillon},
  {Wilson}, {Butterley}, {Yunus}, {Channa}, {Jeffers}, {Falcon}, {Ross},
  {Hestenes}, {Stegman}, {Zhang}, \& {Kumar}}]{VanDyk2019}
{Van Dyk}, S.~D., {Zheng}, W., {Maund}, J.~R., {et~al.} 2019, \apj, 875, 136

\bibitem[{{Woosley} {et~al.}(1994){Woosley}, {Eastman}, {Weaver}, \&
  {Pinto}}]{Woosley1994}
{Woosley}, S.~E., {Eastman}, R.~G., {Weaver}, T.~A., \& {Pinto}, P.~A. 1994,
  \apj, 429, 300

\bibitem[{{Worters} {et~al.}(2016){Worters}, {O'Connor}, {Carter}, {Loubser},
  {Fourie}, {Sickafoose}, \& {Swanevelder}}]{Worters2016}
{Worters}, H.~L., {O'Connor}, J.~E., {Carter}, D.~B., {et~al.} 2016, in Society
  of Photo-Optical Instrumentation Engineers (SPIE) Conference Series, Vol.
  9908, Ground-based and Airborne Instrumentation for Astronomy VI, ed. C.~J.
  {Evans}, L.~{Simard}, \& H.~{Takami}, 99083Y

\bibitem[{{Xiang} {et~al.}(2024){Xiang}, {Mo}, {Wang}, {Wang}, {Zhang}, {Lin},
  {Chen}, {Song}, {Liu}, {Wang}, \& {Li}}]{Xiang2024}
{Xiang}, D., {Mo}, J., {Wang}, X., {et~al.} 2024, \apjl, 969, L15

\bibitem[{{Yaron} \& {Gal-Yam}(2012)}]{WISEREP}
{Yaron}, O. \& {Gal-Yam}, A. 2012, \pasp, 124, 668

\bibitem[{{Yoon} {et~al.}(2010){Yoon}, {Woosley}, \& {Langer}}]{Yoon2010}
{Yoon}, S.~C., {Woosley}, S.~E., \& {Langer}, N. 2010, \apj, 725, 940

\bibitem[{{Zhao} {et~al.}(2025){Zhao}, {Sun}, {Wu}, {Niu}, {Hong}, {Huang},
  {Maund}, {Xi}, {Xiang}, \& {Liu}}]{Zhao2025}
{Zhao}, Y.-H., {Sun}, N.-C., {Wu}, J., {et~al.} 2025, \apjl, 980, L6

\end{thebibliography}

\begin{appendix}
\section{Photometry}

\onecolumn
\begin{longtable}{lccc}
\caption{\SN{} photometry\label{tab:SN_phot}}\\
\hline\hline
Phase (days) & Telescope & Band & AB magnitude\\
\hline
\endfirsthead
\caption{continued.}\\
\hline\hline
Phase (days) & Telescope & Band & AB magnitude\\
\hline
\endhead
\hline
\endfoot
0.02 & ATLAS & $o$ & $19.21\pm0.31$ \\
0.03 & ATLAS & $o$ & $18.64\pm0.16$ \\
1.67 & ATLAS & $o$ & $16.83\pm0.05$ \\
1.68 & ATLAS & $o$ & $16.77\pm0.04$ \\
1.69 & ATLAS & $o$ & $16.73\pm0.04$ \\
1.72 & ATLAS & $o$ & $16.85\pm0.04$ \\
2.49 & Lesedi/Mookodi & $g$ & $17.17\pm0.02$ \\
2.49 & Lesedi/Mookodi & $i$ & $16.79\pm0.04$ \\
2.49 & Lesedi/Mookodi & $r$ & $16.72\pm0.04$ \\
2.57 & \Swift{}/UVOT & $uvw1$ & $20.22\pm0.18$ \\
2.58  & \Swift{}/UVOT & $uvm2$ & $>18.94$ \\ 
2.58 & \Swift{}/UVOT & $B$ & $17.40\pm0.08$ \\
2.58 & \Swift{}/UVOT & $U$ & $18.89\pm0.13$ \\
2.58 & \Swift{}/UVOT & $V$ & $16.92\pm0.10$ \\
2.58 & \Swift{}/UVOT & $uvw2$ & $20.78\pm0.17$ \\
3.27 & \Swift{}/UVOT & $B$ & $17.49\pm0.06$ \\
3.27 & \Swift{}/UVOT & $U$ & $19.12\pm0.10$ \\
3.27 & \Swift{}/UVOT & $V$ & $16.79\pm0.07$ \\
3.27 & \Swift{}/UVOT & $uvw1$ & $19.97\pm0.11$ \\
3.27 & \Swift{}/UVOT & $uvw2$ & $20.85\pm0.13$ \\
3.28 & \Swift{}/UVOT & $uvm2$ & $20.85\pm0.18$ \\
3.62 & Lesedi/Mookodi & $g$ & $17.24\pm0.04$ \\
3.62 & Lesedi/Mookodi & $i$ & $16.59\pm0.03$ \\
3.62 & Lesedi/Mookodi & $r$ & $16.58\pm0.03$ \\
3.62 & Lesedi/Mookodi & $z$ & $16.65\pm0.05$ \\
3.94 & ATLAS & $o$ & $16.42\pm0.03$ \\
3.94 & ATLAS & $o$ & $16.50\pm0.02$ \\
3.95 & ATLAS & $o$ & $16.45\pm0.03$ \\
3.98 & ATLAS & $o$ & $16.48\pm0.03$ \\
4.08 & \Swift{}/UVOT & $B$ & $17.53\pm0.09$ \\
4.08 & \Swift{}/UVOT & $U$ & $18.87\pm0.09$ \\
4.08 & \Swift{}/UVOT & $uvw1$ & $19.64\pm0.09$ \\
4.70 & Lesedi/Mookodi & $g$ & $17.11\pm0.03$ \\
4.70 & Lesedi/Mookodi & $i$ & $16.45\pm0.05$ \\
4.70 & Lesedi/Mookodi & $r$ & $16.44\pm0.05$ \\
4.70 & Lesedi/Mookodi & $z$ & $16.53\pm0.10$ \\
4.99 & \Swift{}/UVOT & $B$ & $17.15\pm0.09$ \\
4.99 & \Swift{}/UVOT & $U$ & $18.58\pm0.14$ \\
4.99 & \Swift{}/UVOT & $V$ & $16.43\pm0.10$ \\
4.99 & \Swift{}/UVOT & $uvm2$ & $20.05\pm0.17$ \\
4.99 & \Swift{}/UVOT & $uvw1$ & $19.23\pm0.16$ \\
4.99 & \Swift{}/UVOT & $uvw2$ & $19.99\pm0.15$ \\
6.18 & \Swift{}/UVOT & $B$ & $16.87\pm0.07$ \\
6.18 & \Swift{}/UVOT & $U$ & $17.84\pm0.09$ \\
6.18 & \Swift{}/UVOT & $V$ & $16.26\pm0.09$ \\
6.18 & \Swift{}/UVOT & $uvw1$ & $18.91\pm0.10$ \\
6.18 & \Swift{}/UVOT & $uvw2$ & $19.94\pm0.13$ \\
6.19 & \Swift{}/UVOT & $uvm2$ & $19.57\pm0.11$ \\
6.50 & Lesedi/Mookodi & $g$ & $16.41\pm0.02$ \\
6.50 & Lesedi/Mookodi & $i$ & $15.87\pm0.04$ \\
6.50 & Lesedi/Mookodi & $r$ & $15.80\pm0.03$ \\
6.50 & Lesedi/Mookodi & $z$ & $15.96\pm0.05$ \\
7.36 & \Swift{}/UVOT & $B$ & $16.31\pm0.06$ \\
7.36 & \Swift{}/UVOT & $U$ & $17.52\pm0.09$ \\
7.36 & \Swift{}/UVOT & $V$ & $15.90\pm0.08$ \\
7.36 & \Swift{}/UVOT & $uvm2$ & $19.87\pm0.15$ \\
7.36 & \Swift{}/UVOT & $uvw1$ & $18.84\pm0.11$ \\
7.36 & \Swift{}/UVOT & $uvw2$ & $19.74\pm0.13$ \\
7.51 & Lesedi/Mookodi & $g$ & $16.02\pm0.02$ \\
7.51 & Lesedi/Mookodi & $i$ & $15.55\pm0.04$ \\
7.51 & Lesedi/Mookodi & $r$ & $15.48\pm0.03$ \\
7.51 & Lesedi/Mookodi & $z$ & $15.67\pm0.05$ \\
7.94 & ATLAS & $o$ & $15.33\pm0.01$ \\
7.98 & ATLAS & $o$ & $15.31\pm0.01$ \\
7.99 & ATLAS & $o$ & $15.32\pm0.01$ \\
8.51 & Lesedi/Mookodi & $g$ & $15.68\pm0.02$ \\
8.51 & Lesedi/Mookodi & $i$ & $15.30\pm0.04$ \\
8.51 & Lesedi/Mookodi & $r$ & $15.19\pm0.04$ \\
8.51 & Lesedi/Mookodi & $z$ & $15.45\pm0.05$ \\
8.60 & \Swift{}/UVOT & $B$ & $15.80\pm0.05$ \\
8.60 & \Swift{}/UVOT & $U$ & $16.96\pm0.06$ \\
8.60 & \Swift{}/UVOT & $uvw1$ & $18.54\pm0.09$ \\
8.61 & \Swift{}/UVOT & $uvw2$ & $19.33\pm0.13$ \\
9.18 & \Swift{}/UVOT & $uvw1$ & $18.43\pm0.08$ \\
9.19 & \Swift{}/UVOT & $B$ & $15.59\pm0.04$ \\
9.19 & \Swift{}/UVOT & $U$ & $16.68\pm0.05$ \\
9.19 & \Swift{}/UVOT & $V$ & $15.22\pm0.06$ \\
9.19 & \Swift{}/UVOT & $uvm2$ & $19.51\pm0.10$ \\
9.19 & \Swift{}/UVOT & $uvw2$ & $19.40\pm0.10$ \\
9.50 & Lesedi/Mookodi & $g$ & $15.40\pm0.02$ \\
9.50 & Lesedi/Mookodi & $i$ & $15.08\pm0.04$ \\
9.50 & Lesedi/Mookodi & $r$ & $14.96\pm0.04$ \\
9.50 & Lesedi/Mookodi & $z$ & $15.27\pm0.06$ \\
9.63 & ATLAS & $o$ & $14.93\pm0.01$ \\
9.63 & ATLAS & $o$ & $14.95\pm0.01$ \\
9.64 & ATLAS & $o$ & $14.92\pm0.01$ \\
9.64 & ATLAS & $o$ & $14.95\pm0.01$ \\
9.67 & ATLAS & $o$ & $14.92\pm0.01$ \\
9.67 & ATLAS & $o$ & $14.92\pm0.01$ \\
9.71 & ATLAS & $o$ & $14.92\pm0.01$ \\
9.71 & ATLAS & $o$ & $14.93\pm0.01$ \\
10.50 & Lesedi/Mookodi & $g$ & $15.15\pm0.02$ \\
10.50 & Lesedi/Mookodi & $i$ & $14.88\pm0.04$ \\
10.50 & Lesedi/Mookodi & $r$ & $14.76\pm0.03$ \\
10.50 & Lesedi/Mookodi & $z$ & $15.05\pm0.05$ \\
10.56 & \Swift{}/UVOT & $B$ & $15.26\pm0.04$ \\
10.56 & \Swift{}/UVOT & $U$ & $16.31\pm0.04$ \\
10.56 & \Swift{}/UVOT & $V$ & $14.98\pm0.05$ \\
10.56 & \Swift{}/UVOT & $uvw1$ & $18.00\pm0.06$ \\
10.56 & \Swift{}/UVOT & $uvw2$ & $18.95\pm0.07$ \\
10.57 & \Swift{}/UVOT & $uvm2$ & $19.11\pm0.08$ \\
11.01 & \Swift{}/UVOT & $B$ & $15.16\pm0.04$ \\
11.01 & \Swift{}/UVOT & $U$ & $16.16\pm0.04$ \\
11.01 & \Swift{}/UVOT & $uvw1$ & $18.05\pm0.07$ \\
11.02 & \Swift{}/UVOT & $V$ & $14.82\pm0.05$ \\
11.02 & \Swift{}/UVOT & $uvm2$ & $18.83\pm0.07$ \\
11.02 & \Swift{}/UVOT & $uvw2$ & $18.76\pm0.07$ \\
14.73 & Lesedi/Mookodi & $g$ & $14.43\pm0.04$ \\
14.73 & Lesedi/Mookodi & $i$ & $14.30\pm0.07$ \\
14.73 & Lesedi/Mookodi & $r$ & $14.16\pm0.04$ \\
14.73 & Lesedi/Mookodi & $z$ & $14.49\pm0.09$ \\
17.63 & ATLAS & $o$ & $13.92\pm0.01$ \\
17.64 & ATLAS & $o$ & $13.92\pm0.01$ \\
17.67 & ATLAS & $o$ & $13.93\pm0.04$ \\
17.91 & \Swift{}/UVOT & $B$ & $14.20\pm0.03$ \\
17.91 & \Swift{}/UVOT & $U$ & $15.08\pm0.03$ \\
17.91 & \Swift{}/UVOT & $uvw1$ & $16.98\pm0.04$ \\
17.91 & \Swift{}/UVOT & $uvw2$ & $18.20\pm0.05$ \\
17.92 & \Swift{}/UVOT & $V$ & $14.01\pm0.03$ \\
17.92 & \Swift{}/UVOT & $uvm2$ & $18.58\pm0.05$ \\
20.56 & Lesedi/Mookodi & $g$ & $14.08\pm0.03$ \\
20.57 & Lesedi/Mookodi & $i$ & $13.84\pm0.03$ \\
20.57 & Lesedi/Mookodi & $r$ & $13.77\pm0.03$ \\
20.57 & Lesedi/Mookodi & $z$ & $14.00\pm0.06$ \\
21.54 & Lesedi/Mookodi & $g$ & $14.09\pm0.03$ \\
21.54 & Lesedi/Mookodi & $i$ & $13.80\pm0.02$ \\
21.54 & Lesedi/Mookodi & $r$ & $13.75\pm0.03$ \\
21.54 & Lesedi/Mookodi & $z$ & $13.98\pm0.04$ \\
21.67 & ATLAS & $o$ & $13.73\pm0.03$ \\
21.70 & ATLAS & $o$ & $13.62\pm0.03$ \\
21.73 & ATLAS & $o$ & $13.71\pm0.01$ \\
21.73 & ATLAS & $o$ & $13.72\pm0.01$ \\
22.45 & \Swift{}/UVOT & $U$ & $15.19\pm0.03$ \\
22.45 & \Swift{}/UVOT & $uvw1$ & $17.24\pm0.04$ \\
22.46 & \Swift{}/UVOT & $B$ & $14.15\pm0.03$ \\
22.46 & \Swift{}/UVOT & $V$ & $13.84\pm0.03$ \\
22.46 & \Swift{}/UVOT & $uvw2$ & $18.57\pm0.06$ \\
22.47 & \Swift{}/UVOT & $uvm2$ & $19.08\pm0.07$ \\
22.56 & Lesedi/Mookodi & $g$ & $14.08\pm0.03$ \\
22.56 & Lesedi/Mookodi & $i$ & $13.77\pm0.04$ \\
22.56 & Lesedi/Mookodi & $r$ & $13.74\pm0.03$ \\
22.56 & Lesedi/Mookodi & $z$ & $13.96\pm0.04$ \\
23.53 & Lesedi/Mookodi & $g$ & $14.11\pm0.03$ \\
23.53 & Lesedi/Mookodi & $i$ & $13.75\pm0.03$ \\
23.53 & Lesedi/Mookodi & $r$ & $13.82\pm0.03$ \\
23.53 & Lesedi/Mookodi & $z$ & $13.93\pm0.05$ \\
24.48 & \Swift{}/UVOT & $B$ & $14.24\pm0.03$ \\
24.48 & \Swift{}/UVOT & $U$ & $15.48\pm0.03$ \\
24.48 & \Swift{}/UVOT & $uvw1$ & $17.53\pm0.04$ \\
24.48 & \Swift{}/UVOT & $uvw2$ & $18.78\pm0.06$ \\
24.49 & \Swift{}/UVOT & $V$ & $13.87\pm0.03$ \\
24.49 & \Swift{}/UVOT & $uvm2$ & $19.28\pm0.07$ \\
24.51 & Lesedi/Mookodi & $g$ & $14.19\pm0.03$ \\
24.51 & Lesedi/Mookodi & $i$ & $13.75\pm0.04$ \\
24.51 & Lesedi/Mookodi & $r$ & $13.74\pm0.03$ \\
24.51 & Lesedi/Mookodi & $z$ & $13.94\pm0.05$ \\
25.51 & Lesedi/Mookodi & $g$ & $14.22\pm0.07$ \\
25.51 & Lesedi/Mookodi & $i$ & $13.73\pm0.05$ \\
25.51 & Lesedi/Mookodi & $r$ & $13.75\pm0.04$ \\
25.51 & Lesedi/Mookodi & $z$ & $13.95\pm0.06$ \\
25.53 & Lesedi/Mookodi & $g$ & $14.21\pm0.05$ \\
25.53 & Lesedi/Mookodi & $i$ & $13.74\pm0.05$ \\
25.53 & Lesedi/Mookodi & $r$ & $13.74\pm0.03$ \\
25.53 & Lesedi/Mookodi & $z$ & $13.92\pm0.05$ \\
25.63 & ATLAS & $o$ & $13.70\pm0.00$ \\
25.63 & ATLAS & $o$ & $13.71\pm0.00$ \\
25.63 & ATLAS & $o$ & $13.73\pm0.00$ \\
25.64 & ATLAS & $o$ & $13.71\pm0.00$ \\
25.64 & ATLAS & $o$ & $13.71\pm0.00$ \\
25.64 & ATLAS & $o$ & $13.74\pm0.00$ \\
25.65 & ATLAS & $o$ & $13.72\pm0.00$ \\
25.66 & ATLAS & $o$ & $13.70\pm0.00$ \\
26.41 & \Swift{}/UVOT & $B$ & $14.43\pm0.03$ \\
26.41 & \Swift{}/UVOT & $U$ & $15.84\pm0.03$ \\
26.41 & \Swift{}/UVOT & $uvw1$ & $17.81\pm0.05$ \\
26.42 & \Swift{}/UVOT & $V$ & $13.95\pm0.03$ \\
26.42 & \Swift{}/UVOT & $uvm2$ & $19.56\pm0.07$ \\
26.42 & \Swift{}/UVOT & $uvw2$ & $19.13\pm0.06$ \\
26.51 & Lesedi/Mookodi & $g$ & $14.32\pm0.03$ \\
26.51 & Lesedi/Mookodi & $i$ & $13.73\pm0.04$ \\
26.51 & Lesedi/Mookodi & $r$ & $13.76\pm0.03$ \\
26.51 & Lesedi/Mookodi & $z$ & $13.94\pm0.06$ \\
27.51 & Lesedi/Mookodi & $g$ & $14.41\pm0.03$ \\
27.51 & Lesedi/Mookodi & $i$ & $13.74\pm0.05$ \\
27.51 & Lesedi/Mookodi & $r$ & $13.79\pm0.03$ \\
27.51 & Lesedi/Mookodi & $z$ & $13.93\pm0.04$ \\
27.88 & ATLAS & $o$ & $13.69\pm0.00$ \\
27.88 & ATLAS & $o$ & $13.69\pm0.00$ \\
27.88 & ATLAS & $o$ & $13.70\pm0.00$ \\
27.92 & ATLAS & $o$ & $13.70\pm0.00$ \\
28.52 & Lesedi/Mookodi & $g$ & $14.50\pm0.03$ \\
28.52 & Lesedi/Mookodi & $i$ & $13.76\pm0.03$ \\
28.52 & Lesedi/Mookodi & $r$ & $13.84\pm0.03$ \\
28.53 & Lesedi/Mookodi & $z$ & $13.93\pm0.05$ \\
29.62 & ATLAS & $o$ & $13.77\pm0.00$ \\
29.62 & ATLAS & $o$ & $13.77\pm0.00$ \\
29.63 & ATLAS & $o$ & $13.76\pm0.00$ \\
29.66 & ATLAS & $o$ & $13.75\pm0.00$ \\
30.56 & Lesedi/Mookodi & $g$ & $14.66\pm0.10$ \\
30.56 & Lesedi/Mookodi & $i$ & $13.80\pm0.05$ \\
30.56 & Lesedi/Mookodi & $r$ & $13.90\pm0.04$ \\
30.56 & Lesedi/Mookodi & $z$ & $13.97\pm0.05$ \\
31.59 & Lesedi/Mookodi & $g$ & $14.83\pm0.04$ \\
31.59 & Lesedi/Mookodi & $i$ & $13.85\pm0.03$ \\
31.59 & Lesedi/Mookodi & $r$ & $14.00\pm0.03$ \\
31.86 & ATLAS & $o$ & $13.83\pm0.01$ \\
31.86 & ATLAS & $o$ & $13.84\pm0.00$ \\
31.87 & ATLAS & $o$ & $13.88\pm0.01$ \\
31.88 & ATLAS & $o$ & $13.84\pm0.00$ \\
32.51 & Lesedi/Mookodi & $g$ & $14.92\pm0.02$ \\
32.51 & Lesedi/Mookodi & $i$ & $13.87\pm0.04$ \\
32.51 & Lesedi/Mookodi & $r$ & $14.05\pm0.04$ \\
32.51 & Lesedi/Mookodi & $z$ & $14.03\pm0.04$ \\
33.52 & Lesedi/Mookodi & $g$ & $15.01\pm0.03$ \\
33.52 & \Swift{}/UVOT & $uvw1$ & $18.97\pm0.10$ \\
33.53 & Lesedi/Mookodi & $i$ & $13.90\pm0.03$ \\
33.53 & Lesedi/Mookodi & $r$ & $14.09\pm0.03$ \\
33.53 & Lesedi/Mookodi & $z$ & $14.03\pm0.04$ \\
33.62 & ATLAS & $o$ & $13.94\pm0.00$ \\
33.63 & ATLAS & $o$ & $13.94\pm0.00$ \\
33.63 & ATLAS & $o$ & $13.95\pm0.00$ \\
33.64 & ATLAS & $o$ & $13.93\pm0.00$ \\
34.55 & Lesedi/Mookodi & $g$ & $15.10\pm0.02$ \\
34.55 & Lesedi/Mookodi & $i$ & $13.95\pm0.03$ \\
34.55 & Lesedi/Mookodi & $r$ & $14.15\pm0.03$ \\
34.55 & Lesedi/Mookodi & $z$ & $14.07\pm0.05$ \\
35.87 & ATLAS & $o$ & $13.88\pm0.00$ \\
35.88 & ATLAS & $o$ & $13.94\pm0.00$ \\
35.88 & ATLAS & $o$ & $14.03\pm0.00$ \\
35.89 & ATLAS & $o$ & $13.88\pm0.00$ \\
35.93 & \Swift{}/UVOT & $B$ & $15.42\pm0.04$ \\
35.93 & \Swift{}/UVOT & $U$ & $17.37\pm0.07$ \\
35.93 & \Swift{}/UVOT & $V$ & $14.54\pm0.04$ \\
35.93 & \Swift{}/UVOT & $uvw1$ & $19.20\pm0.12$ \\
35.93 & \Swift{}/UVOT & $uvw2$ & $20.34\pm0.17$ \\
35.94 & \Swift{}/UVOT & $uvm2$ & $20.84\pm0.22$ \\
37.65 & Lesedi/Mookodi & $g$ & $15.30\pm0.03$ \\
37.66 & Lesedi/Mookodi & $i$ & $14.03\pm0.03$ \\
37.66 & Lesedi/Mookodi & $r$ & $14.28\pm0.03$ \\
37.66 & Lesedi/Mookodi & $z$ & $14.11\pm0.06$ \\
39.07 & \Swift{}/UVOT & $B$ & $15.73\pm0.04$ \\
39.07 & \Swift{}/UVOT & $U$ & $17.81\pm0.07$ \\
39.07 & \Swift{}/UVOT & $V$ & $14.74\pm0.04$ \\
39.07 & \Swift{}/UVOT & $uvw1$ & $19.60\pm0.12$ \\
39.07 & \Swift{}/UVOT & $uvw2$ & $20.35\pm0.13$ \\
39.08 & \Swift{}/UVOT & $uvm2$ & $20.83\pm0.18$ \\
39.85 & ATLAS & $o$ & $14.15\pm0.01$ \\
39.86 & ATLAS & $o$ & $14.11\pm0.01$ \\
39.86 & ATLAS & $o$ & $14.13\pm0.01$ \\
39.87 & ATLAS & $o$ & $14.12\pm0.01$ \\
40.53 & Lesedi/Mookodi & $g$ & $15.43\pm0.03$ \\
40.53 & Lesedi/Mookodi & $i$ & $14.12\pm0.03$ \\
40.53 & Lesedi/Mookodi & $r$ & $14.41\pm0.03$ \\
40.53 & Lesedi/Mookodi & $z$ & $14.17\pm0.05$ \\
41.64 & Lesedi/Mookodi & $g$ & $15.47\pm0.07$ \\
41.64 & Lesedi/Mookodi & $i$ & $14.17\pm0.13$ \\
41.64 & Lesedi/Mookodi & $r$ & $14.46\pm0.14$ \\
41.64 & Lesedi/Mookodi & $z$ & $14.25\pm0.10$ \\
42.43 & \Swift{}/UVOT & $uvw1$ & $19.60\pm0.13$ \\
42.44 & \Swift{}/UVOT & $B$ & $15.85\pm0.04$ \\
42.44 & \Swift{}/UVOT & $U$ & $17.95\pm0.08$ \\
42.44 & \Swift{}/UVOT & $V$ & $14.83\pm0.04$ \\
42.44 & \Swift{}/UVOT & $uvm2$ & $21.17\pm0.24$ \\
42.44 & \Swift{}/UVOT & $uvw2$ & $20.54\pm0.16$ \\
42.67 & Lesedi/Mookodi & $g$ & $15.53\pm0.03$ \\
42.67 & Lesedi/Mookodi & $i$ & $14.21\pm0.03$ \\
42.67 & Lesedi/Mookodi & $r$ & $14.50\pm0.03$ \\
42.67 & Lesedi/Mookodi & $z$ & $14.25\pm0.08$ \\
43.86 & ATLAS & $o$ & $14.41\pm0.01$ \\
43.87 & ATLAS & $o$ & $14.40\pm0.01$ \\
43.87 & ATLAS & $o$ & $14.41\pm0.01$ \\
43.88 & ATLAS & $o$ & $14.39\pm0.01$ \\
44.69 & Lesedi/Mookodi & $g$ & $15.63\pm0.04$ \\
44.69 & Lesedi/Mookodi & $i$ & $14.30\pm0.02$ \\
44.69 & Lesedi/Mookodi & $r$ & $14.60\pm0.01$ \\
44.69 & Lesedi/Mookodi & $z$ & $14.26\pm0.05$ \\
45.57 & \Swift{}/UVOT & $B$ & $15.96\pm0.04$ \\
45.57 & \Swift{}/UVOT & $U$ & $17.98\pm0.07$ \\
45.57 & \Swift{}/UVOT & $uvw1$ & $19.45\pm0.11$ \\
45.57 & \Swift{}/UVOT & $uvw2$ & $20.26\pm0.11$ \\
45.65 & Lesedi/Mookodi & $g$ & $15.66\pm0.03$ \\
45.65 & Lesedi/Mookodi & $i$ & $14.33\pm0.03$ \\
45.65 & Lesedi/Mookodi & $r$ & $14.64\pm0.03$ \\
45.65 & Lesedi/Mookodi & $z$ & $14.32\pm0.06$ \\
46.06 & \Swift{}/UVOT & $V$ & $15.00\pm0.05$ \\
46.06 & \Swift{}/UVOT & $uvm2$ & $21.03\pm0.31$ \\
46.51 & Lesedi/Mookodi & $g$ & $15.67\pm0.03$ \\
46.51 & Lesedi/Mookodi & $i$ & $14.36\pm0.02$ \\
46.51 & Lesedi/Mookodi & $r$ & $14.68\pm0.03$ \\
46.51 & Lesedi/Mookodi & $z$ & $14.30\pm0.04$ \\
47.51 & Lesedi/Mookodi & $g$ & $15.69\pm0.05$ \\
47.51 & Lesedi/Mookodi & $i$ & $14.38\pm0.03$ \\
47.51 & Lesedi/Mookodi & $r$ & $14.70\pm0.03$ \\
47.51 & Lesedi/Mookodi & $z$ & $14.33\pm0.04$ \\
47.83 & ATLAS & $o$ & $14.33\pm0.01$ \\
47.84 & ATLAS & $o$ & $14.28\pm0.01$ \\
47.84 & ATLAS & $o$ & $14.31\pm0.01$ \\
47.85 & ATLAS & $o$ & $14.31\pm0.01$ \\
49.23 & \Swift{}/UVOT & $U$ & $18.06\pm0.09$ \\
49.23 & \Swift{}/UVOT & $uvw1$ & $19.64\pm0.12$ \\
49.47 & \Swift{}/UVOT & $B$ & $16.10\pm0.07$ \\
49.47 & \Swift{}/UVOT & $V$ & $15.13\pm0.07$ \\
49.47 & \Swift{}/UVOT & $uvm2$ & $20.81\pm0.31$ \\
49.47 & \Swift{}/UVOT & $uvw2$ & $20.38\pm0.25$ \\
50.58 & Lesedi/Mookodi & $g$ & $15.81\pm0.05$ \\
50.58 & Lesedi/Mookodi & $i$ & $14.49\pm0.03$ \\
50.58 & Lesedi/Mookodi & $r$ & $14.80\pm0.03$ \\
50.58 & Lesedi/Mookodi & $z$ & $14.41\pm0.05$ \\
51.84 & ATLAS & $o$ & $14.65\pm0.01$ \\
51.85 & ATLAS & $o$ & $14.63\pm0.01$ \\
51.86 & ATLAS & $o$ & $14.63\pm0.01$ \\
51.87 & ATLAS & $o$ & $14.61\pm0.01$ \\
52.51 & Lesedi/Mookodi & $g$ & $15.81\pm0.03$ \\
52.51 & Lesedi/Mookodi & $i$ & $14.54\pm0.03$ \\
52.51 & Lesedi/Mookodi & $r$ & $14.86\pm0.03$ \\
52.51 & Lesedi/Mookodi & $z$ & $14.45\pm0.04$ \\
53.36 & \Swift{}/UVOT & $uvw1$ & $19.76\pm0.21$ \\
53.37 & \Swift{}/UVOT & $B$ & $16.25\pm0.05$ \\
53.37 & \Swift{}/UVOT & $U$ & $18.20\pm0.11$ \\
53.37 & \Swift{}/UVOT & $V$ & $15.16\pm0.05$ \\
53.37 & \Swift{}/UVOT & $uvw2$ & $20.62\pm0.21$ \\
53.38 & \Swift{}/UVOT & $uvm2$ & $20.91\pm0.22$ \\
53.58 & ATLAS & $o$ & $14.73\pm0.01$ \\
53.58 & ATLAS & $o$ & $14.75\pm0.01$ \\
53.59 & ATLAS & $o$ & $14.75\pm0.01$ \\
53.60 & ATLAS & $o$ & $14.75\pm0.01$ \\
53.65 & Lesedi/Mookodi & $g$ & $15.88\pm0.03$ \\
53.65 & Lesedi/Mookodi & $i$ & $14.60\pm0.04$ \\
53.65 & Lesedi/Mookodi & $r$ & $14.89\pm0.03$ \\
53.65 & Lesedi/Mookodi & $z$ & $14.49\pm0.06$ \\
54.67 & Lesedi/Mookodi & $g$ & $15.89\pm0.04$ \\
54.67 & Lesedi/Mookodi & $i$ & $14.61\pm0.03$ \\
54.67 & Lesedi/Mookodi & $r$ & $14.91\pm0.02$ \\
54.67 & Lesedi/Mookodi & $z$ & $14.50\pm0.06$ \\
55.84 & ATLAS & $o$ & $14.77\pm0.01$ \\
55.84 & ATLAS & $o$ & $14.77\pm0.01$ \\
55.84 & ATLAS & $o$ & $14.80\pm0.01$ \\
55.87 & ATLAS & $o$ & $14.78\pm0.01$ \\
56.58 & Lesedi/Mookodi & $g$ & $15.90\pm0.03$ \\
56.58 & Lesedi/Mookodi & $i$ & $14.65\pm0.03$ \\
56.58 & Lesedi/Mookodi & $r$ & $14.96\pm0.03$ \\
56.58 & Lesedi/Mookodi & $z$ & $14.54\pm0.07$ \\
57.17 & \Swift{}/UVOT & $B$ & $16.08\pm0.07$ \\
57.17 & \Swift{}/UVOT & $U$ & $18.13\pm0.15$ \\
57.17 & \Swift{}/UVOT & $V$ & $15.21\pm0.08$ \\
57.17 & \Swift{}/UVOT & $uvw1$ & $19.49\pm0.20$ \\
57.17 & \Swift{}/UVOT & $uvw2$ & $20.57\pm0.28$ \\
57.18 & \Swift{}/UVOT & $uvm2$ & $20.74\pm0.32$ \\
57.57 & ATLAS & $o$ & $14.84\pm0.01$ \\
57.57 & ATLAS & $o$ & $14.85\pm0.01$ \\
57.58 & ATLAS & $o$ & $14.80\pm0.01$ \\
57.59 & ATLAS & $o$ & $14.84\pm0.01$ \\
57.67 & Lesedi/Mookodi & $g$ & $15.89\pm0.06$ \\
57.67 & Lesedi/Mookodi & $i$ & $14.67\pm0.03$ \\
57.67 & Lesedi/Mookodi & $r$ & $14.97\pm0.02$ \\
57.67 & Lesedi/Mookodi & $z$ & $14.54\pm0.05$ \\
59.56 & Lesedi/Mookodi & $i$ & $14.71\pm0.02$ \\
59.56 & Lesedi/Mookodi & $r$ & $15.03\pm0.03$ \\
59.56 & Lesedi/Mookodi & $z$ & $14.59\pm0.06$ \\
59.58 & Lesedi/Mookodi & $g$ & $15.96\pm0.03$ \\
59.86 & ATLAS & $o$ & $14.87\pm0.01$ \\
59.86 & ATLAS & $o$ & $14.89\pm0.01$ \\
59.87 & ATLAS & $o$ & $14.80\pm0.01$ \\
59.87 & ATLAS & $o$ & $14.85\pm0.01$ \\
59.87 & ATLAS & $o$ & $14.88\pm0.01$ \\
59.88 & ATLAS & $o$ & $14.57\pm0.01$ \\
59.89 & ATLAS & $o$ & $14.87\pm0.01$ \\
59.90 & ATLAS & $o$ & $14.86\pm0.01$ \\
60.65 & Lesedi/Mookodi & $g$ & $15.94\pm0.09$ \\
60.65 & Lesedi/Mookodi & $i$ & $14.73\pm0.02$ \\
60.65 & Lesedi/Mookodi & $r$ & $15.04\pm0.02$ \\
60.65 & Lesedi/Mookodi & $z$ & $14.59\pm0.06$ \\
61.52 & Lesedi/Mookodi & $g$ & $15.97\pm0.03$ \\
61.52 & Lesedi/Mookodi & $i$ & $14.75\pm0.04$ \\
61.52 & Lesedi/Mookodi & $r$ & $15.06\pm0.04$ \\
61.52 & Lesedi/Mookodi & $z$ & $14.60\pm0.07$ \\
62.57 & Lesedi/Mookodi & $g$ & $16.01\pm0.03$ \\
62.57 & Lesedi/Mookodi & $i$ & $14.77\pm0.03$ \\
62.57 & Lesedi/Mookodi & $r$ & $15.08\pm0.04$ \\
62.57 & Lesedi/Mookodi & $z$ & $14.63\pm0.05$ \\
63.10 & \Swift{}/UVOT & $B$ & $16.27\pm0.05$ \\
63.10 & \Swift{}/UVOT & $U$ & $17.99\pm0.08$ \\
63.10 & \Swift{}/UVOT & $uvw1$ & $19.92\pm0.17$ \\
63.11 & \Swift{}/UVOT & $V$ & $15.43\pm0.05$ \\
63.11 & \Swift{}/UVOT & $uvm2$ & $21.42\pm0.30$ \\
63.11 & \Swift{}/UVOT & $uvw2$ & $20.81\pm0.19$ \\
63.57 & Lesedi/Mookodi & $g$ & $15.99\pm0.04$ \\
63.57 & Lesedi/Mookodi & $i$ & $14.80\pm0.04$ \\
63.57 & Lesedi/Mookodi & $r$ & $15.10\pm0.04$ \\
63.57 & Lesedi/Mookodi & $z$ & $14.66\pm0.08$ \\
63.88 & ATLAS & $o$ & $14.96\pm0.01$ \\
63.88 & ATLAS & $o$ & $14.97\pm0.01$ \\
63.89 & ATLAS & $o$ & $14.97\pm0.01$ \\
63.92 & ATLAS & $o$ & $14.94\pm0.01$ \\
65.55 & ATLAS & $o$ & $15.00\pm0.01$ \\
65.55 & ATLAS & $o$ & $15.00\pm0.01$ \\
65.56 & ATLAS & $o$ & $15.00\pm0.01$ \\
65.57 & ATLAS & $o$ & $15.00\pm0.01$ \\
65.67 & Lesedi/Mookodi & $g$ & $15.99\pm0.05$ \\
65.67 & Lesedi/Mookodi & $r$ & $15.12\pm0.03$ \\
65.68 & Lesedi/Mookodi & $i$ & $14.82\pm0.03$ \\
65.68 & Lesedi/Mookodi & $z$ & $14.66\pm0.06$ \\
67.55 & Lesedi/Mookodi & $g$ & $16.05\pm0.03$ \\
67.55 & Lesedi/Mookodi & $i$ & $14.87\pm0.02$ \\
67.55 & Lesedi/Mookodi & $r$ & $15.17\pm0.03$ \\
67.55 & Lesedi/Mookodi & $z$ & $14.69\pm0.06$ \\
68.20 & \Swift{}/UVOT & $B$ & $16.35\pm0.05$ \\
68.20 & \Swift{}/UVOT & $U$ & $18.14\pm0.09$ \\
68.20 & \Swift{}/UVOT & $V$ & $15.53\pm0.06$ \\
68.20 & \Swift{}/UVOT & $uvm2$ & $20.96\pm0.21$ \\
68.20 & \Swift{}/UVOT & $uvw1$ & $20.03\pm0.21$ \\
68.20 & \Swift{}/UVOT & $uvw2$ & $20.74\pm0.20$ \\
69.54 & ATLAS & $o$ & $15.05\pm0.01$ \\
69.54 & ATLAS & $o$ & $15.08\pm0.01$ \\
69.55 & ATLAS & $o$ & $15.04\pm0.01$ \\
69.56 & ATLAS & $o$ & $15.07\pm0.01$ \\
69.56 & Lesedi/Mookodi & $g$ & $16.07\pm0.02$ \\
69.56 & Lesedi/Mookodi & $r$ & $15.20\pm0.03$ \\
69.57 & Lesedi/Mookodi & $g$ & $16.06\pm0.02$ \\
69.57 & Lesedi/Mookodi & $i$ & $14.90\pm0.03$ \\
69.57 & Lesedi/Mookodi & $i$ & $14.92\pm0.03$ \\
69.57 & Lesedi/Mookodi & $r$ & $15.20\pm0.03$ \\
69.57 & Lesedi/Mookodi & $z$ & $14.75\pm0.07$ \\
69.57 & Lesedi/Mookodi & $z$ & $14.75\pm0.07$ \\
71.59 & Lesedi/Mookodi & $g$ & $16.13\pm0.06$ \\
71.59 & Lesedi/Mookodi & $i$ & $14.94\pm0.03$ \\
71.59 & Lesedi/Mookodi & $r$ & $15.23\pm0.03$ \\
71.59 & Lesedi/Mookodi & $z$ & $14.74\pm0.06$ \\
73.15 & \Swift{}/UVOT & $B$ & $16.34\pm0.05$ \\
73.15 & \Swift{}/UVOT & $U$ & $17.83\pm0.09$ \\
73.15 & \Swift{}/UVOT & $uvw1$ & $19.84\pm0.21$ \\
73.16 & \Swift{}/UVOT & $V$ & $15.55\pm0.05$ \\
73.16 & \Swift{}/UVOT & $uvm2$ & $20.89\pm0.19$ \\
73.16 & \Swift{}/UVOT & $uvw2$ & $20.47\pm0.17$ \\
73.52 & Lesedi/Mookodi & $g$ & $16.13\pm0.03$ \\
73.53 & Lesedi/Mookodi & $i$ & $15.01\pm0.03$ \\
73.53 & Lesedi/Mookodi & $r$ & $15.28\pm0.03$ \\
73.53 & Lesedi/Mookodi & $z$ & $14.77\pm0.05$ \\
73.55 & ATLAS & $o$ & $15.16\pm0.01$ \\
73.56 & ATLAS & $o$ & $15.15\pm0.01$ \\
73.56 & ATLAS & $o$ & $15.17\pm0.01$ \\
73.58 & ATLAS & $o$ & $15.17\pm0.01$ \\
77.53 & ATLAS & $o$ & $15.21\pm0.01$ \\
77.53 & ATLAS & $o$ & $15.23\pm0.01$ \\
77.54 & ATLAS & $o$ & $15.22\pm0.01$ \\
77.55 & ATLAS & $o$ & $15.23\pm0.01$ \\
77.89 & \Swift{}/UVOT & $uvw1$ & $19.61\pm0.25$ \\
78.05 & \Swift{}/UVOT & $B$ & $16.41\pm0.06$ \\
78.05 & \Swift{}/UVOT & $U$ & $18.08\pm0.10$ \\
78.05 & \Swift{}/UVOT & $uvw2$ & $20.53\pm0.28$ \\
78.54 & Lesedi/Mookodi & $g$ & $16.16\pm0.03$ \\
78.54 & Lesedi/Mookodi & $i$ & $15.07\pm0.02$ \\
78.54 & Lesedi/Mookodi & $r$ & $15.35\pm0.03$ \\
78.54 & Lesedi/Mookodi & $z$ & $14.87\pm0.06$ \\
80.50 & Lesedi/Mookodi & $g$ & $16.17\pm0.03$ \\
80.50 & Lesedi/Mookodi & $i$ & $15.10\pm0.03$ \\
80.50 & Lesedi/Mookodi & $r$ & $15.37\pm0.03$ \\
80.50 & Lesedi/Mookodi & $z$ & $14.92\pm0.08$ \\
81.53 & ATLAS & $o$ & $15.28\pm0.01$ \\
81.53 & ATLAS & $o$ & $15.29\pm0.01$ \\
81.54 & ATLAS & $o$ & $15.29\pm0.01$ \\
81.55 & ATLAS & $o$ & $15.32\pm0.01$ \\
82.62 & Lesedi/Mookodi & $g$ & $16.19\pm0.03$ \\
82.62 & Lesedi/Mookodi & $i$ & $15.16\pm0.02$ \\
82.62 & Lesedi/Mookodi & $r$ & $15.41\pm0.02$ \\
82.62 & Lesedi/Mookodi & $z$ & $14.94\pm0.06$ \\
83.78 & ATLAS & $o$ & $15.30\pm0.01$ \\
83.78 & ATLAS & $o$ & $15.32\pm0.01$ \\
83.79 & ATLAS & $o$ & $15.34\pm0.01$ \\
83.80 & ATLAS & $o$ & $15.32\pm0.01$ \\
84.32 & \Swift{}/UVOT & $U$ & $18.01\pm0.08$ \\
84.32 & \Swift{}/UVOT & $uvw1$ & $19.93\pm0.16$ \\
84.33 & \Swift{}/UVOT & $B$ & $16.44\pm0.05$ \\
84.33 & \Swift{}/UVOT & $V$ & $15.79\pm0.07$ \\
84.33 & \Swift{}/UVOT & $uvw2$ & $20.87\pm0.20$ \\
85.53 & ATLAS & $o$ & $15.39\pm0.01$ \\
85.54 & ATLAS & $o$ & $15.38\pm0.01$ \\
85.56 & ATLAS & $o$ & $15.37\pm0.01$ \\
85.56 & ATLAS & $o$ & $15.40\pm0.01$ \\
87.81 & ATLAS & $o$ & $15.41\pm0.01$ \\
87.81 & ATLAS & $o$ & $15.42\pm0.01$ \\
87.82 & ATLAS & $o$ & $15.42\pm0.01$ \\
87.84 & ATLAS & $o$ & $15.41\pm0.01$ \\
88.50 & Lesedi/Mookodi & $g$ & $16.24\pm0.06$ \\
88.50 & Lesedi/Mookodi & $r$ & $15.50\pm0.04$ \\
88.51 & Lesedi/Mookodi & $i$ & $15.27\pm0.04$ \\
88.51 & Lesedi/Mookodi & $z$ & $15.06\pm0.08$ \\
89.09 & \Swift{}/UVOT & $B$ & $16.43\pm0.08$ \\
89.09 & \Swift{}/UVOT & $U$ & $17.84\pm0.12$ \\
89.09 & \Swift{}/UVOT & $V$ & $15.81\pm0.09$ \\
89.09 & \Swift{}/UVOT & $uvm2$ & $>20.98$ \\ 
89.09 & \Swift{}/UVOT & $uvw1$ & $19.41\pm0.20$ \\
89.09 & \Swift{}/UVOT & $uvw2$ & $>21.08$ \\
89.51 & ATLAS & $o$ & $15.41\pm0.01$ \\
89.51 & ATLAS & $o$ & $15.43\pm0.01$ \\
89.52 & ATLAS & $o$ & $15.44\pm0.01$ \\
89.53 & ATLAS & $o$ & $15.45\pm0.01$ \\
91.75 & ATLAS & $o$ & $15.49\pm0.01$ \\
91.76 & ATLAS & $o$ & $15.48\pm0.01$ \\
91.77 & ATLAS & $o$ & $15.47\pm0.01$ \\
91.78 & ATLAS & $o$ & $15.52\pm0.01$ \\
92.59 & Lesedi/Mookodi & $g$ & $16.28\pm0.03$ \\
92.59 & Lesedi/Mookodi & $i$ & $15.36\pm0.03$ \\
92.59 & Lesedi/Mookodi & $r$ & $15.56\pm0.02$ \\
92.59 & Lesedi/Mookodi & $z$ & $15.12\pm0.08$ \\
93.52 & ATLAS & $o$ & $15.52\pm0.01$ \\
93.52 & ATLAS & $o$ & $15.52\pm0.01$ \\
93.53 & ATLAS & $o$ & $15.51\pm0.01$ \\
93.55 & ATLAS & $o$ & $15.49\pm0.01$ \\
95.78 & ATLAS & $o$ & $15.52\pm0.01$ \\
95.78 & ATLAS & $o$ & $15.55\pm0.01$ \\
95.78 & ATLAS & $o$ & $15.55\pm0.01$ \\
95.80 & ATLAS & $o$ & $15.54\pm0.01$ \\
99.74 & ATLAS & $o$ & $15.60\pm0.01$ \\
99.74 & ATLAS & $o$ & $15.61\pm0.01$ \\
99.75 & ATLAS & $o$ & $15.60\pm0.01$ \\
99.76 & ATLAS & $o$ & $15.60\pm0.01$ \\
100.57 & Lesedi/Mookodi & $g$ & $16.37\pm0.04$ \\
100.57 & Lesedi/Mookodi & $i$ & $15.51\pm0.03$ \\
100.57 & Lesedi/Mookodi & $r$ & $15.67\pm0.03$ \\
100.57 & Lesedi/Mookodi & $z$ & $15.22\pm0.08$ \\
107.78 & ATLAS & $o$ & $15.64\pm0.01$ \\
107.78 & ATLAS & $o$ & $15.68\pm0.01$ \\
107.80 & ATLAS & $o$ & $15.64\pm0.01$ \\
107.81 & ATLAS & $o$ & $15.65\pm0.01$ \\
111.73 & ATLAS & $o$ & $15.77\pm0.01$ \\
111.74 & ATLAS & $o$ & $15.75\pm0.01$ \\
111.74 & ATLAS & $o$ & $15.81\pm0.01$ \\
111.75 & ATLAS & $o$ & $15.82\pm0.01$ \\
115.80 & ATLAS & $o$ & $15.76\pm0.02$ \\
115.80 & ATLAS & $o$ & $15.78\pm0.02$ \\
115.82 & ATLAS & $o$ & $15.72\pm0.02$ \\
115.84 & ATLAS & $o$ & $15.78\pm0.02$ \\
116.50 & Lesedi/Mookodi & $g$ & $16.53\pm0.04$ \\
116.50 & Lesedi/Mookodi & $i$ & $15.82\pm0.02$ \\
116.50 & Lesedi/Mookodi & $r$ & $15.89\pm0.03$ \\
116.50 & Lesedi/Mookodi & $z$ & $15.52\pm0.06$ \\
117.49 & ATLAS & $o$ & $15.93\pm0.01$ \\
117.50 & ATLAS & $o$ & $15.93\pm0.02$ \\
117.52 & ATLAS & $o$ & $15.93\pm0.02$ \\
117.52 & ATLAS & $o$ & $15.94\pm0.02$ \\
119.75 & ATLAS & $o$ & $15.91\pm0.03$ \\
119.76 & ATLAS & $o$ & $15.79\pm0.03$ \\
119.78 & ATLAS & $o$ & $15.83\pm0.03$ \\
119.78 & ATLAS & $o$ & $15.85\pm0.03$ \\
123.73 & ATLAS & $o$ & $15.86\pm0.02$ \\
123.74 & ATLAS & $o$ & $15.88\pm0.01$ \\
123.75 & ATLAS & $o$ & $15.86\pm0.02$ \\
123.76 & ATLAS & $o$ & $15.84\pm0.02$ \\
124.79 & ATLAS & $o$ & $15.99\pm0.02$ \\
124.82 & ATLAS & $o$ & $15.96\pm0.02$ \\
124.82 & ATLAS & $o$ & $15.99\pm0.02$ \\
124.83 & ATLAS & $o$ & $15.98\pm0.03$ \\
125.48 & ATLAS & $o$ & $16.12\pm0.03$ \\
125.49 & ATLAS & $o$ & $16.09\pm0.02$ \\
125.50 & ATLAS & $o$ & $15.99\pm0.01$ \\
125.51 & ATLAS & $o$ & $16.02\pm0.02$ \\
125.51 & ATLAS & $o$ & $16.06\pm0.01$ \\
125.51 & ATLAS & $o$ & $16.07\pm0.01$ \\
131.49 & Lesedi/Mookodi & $g$ & $16.73\pm0.03$ \\
131.49 & Lesedi/Mookodi & $i$ & $16.12\pm0.02$ \\
131.49 & Lesedi/Mookodi & $r$ & $16.12\pm0.02$ \\
131.49 & Lesedi/Mookodi & $z$ & $15.83\pm0.05$ \\
133.49 & Lesedi/Mookodi & $g$ & $16.74\pm0.04$ \\
133.49 & Lesedi/Mookodi & $i$ & $16.15\pm0.02$ \\
133.49 & Lesedi/Mookodi & $r$ & $16.12\pm0.02$ \\
133.49 & Lesedi/Mookodi & $z$ & $15.86\pm0.06$ \\
133.52 & ATLAS & $o$ & $16.17\pm0.02$ \\
133.53 & ATLAS & $o$ & $16.17\pm0.02$ \\
133.53 & ATLAS & $o$ & $16.19\pm0.02$ \\
133.56 & ATLAS & $o$ & $16.14\pm0.02$ \\
137.51 & ATLAS & $o$ & $16.19\pm0.02$ \\
137.51 & ATLAS & $o$ & $16.26\pm0.02$ \\
137.52 & ATLAS & $o$ & $16.20\pm0.02$ \\
137.54 & ATLAS & $o$ & $16.23\pm0.02$ \\
141.48 & ATLAS & $o$ & $16.33\pm0.02$ \\
141.50 & ATLAS & $o$ & $16.28\pm0.02$ \\
141.51 & ATLAS & $o$ & $16.28\pm0.02$ \\
141.51 & ATLAS & $o$ & $16.28\pm0.02$ \\
143.70 & ATLAS & $o$ & $16.34\pm0.03$ \\
143.70 & ATLAS & $o$ & $16.43\pm0.06$ \\
143.71 & ATLAS & $o$ & $16.31\pm0.03$ \\
143.72 & ATLAS & $o$ & $16.31\pm0.02$ \\
148.44 & Lesedi/Mookodi & $g$ & $16.91\pm0.05$ \\
148.44 & Lesedi/Mookodi & $i$ & $16.42\pm0.03$ \\
148.44 & Lesedi/Mookodi & $r$ & $16.30\pm0.03$ \\
148.44 & Lesedi/Mookodi & $z$ & $16.20\pm0.09$ \\
151.69 & ATLAS & $o$ & $16.38\pm0.05$ \\
151.70 & ATLAS & $o$ & $16.28\pm0.03$ \\
151.71 & ATLAS & $o$ & $16.47\pm0.03$ \\
151.72 & ATLAS & $o$ & $16.32\pm0.04$ \\
153.43 & Lesedi/Mookodi & $g$ & $16.96\pm0.05$ \\
153.44 & Lesedi/Mookodi & $i$ & $16.48\pm0.04$ \\
153.44 & Lesedi/Mookodi & $r$ & $16.36\pm0.03$ \\
153.44 & Lesedi/Mookodi & $z$ & $16.24\pm0.07$ \\
153.45 & ATLAS & $o$ & $16.48\pm0.02$ \\
153.45 & ATLAS & $o$ & $16.50\pm0.02$ \\
153.47 & ATLAS & $o$ & $16.48\pm0.02$ \\
153.48 & ATLAS & $o$ & $16.61\pm0.03$ \\
155.72 & ATLAS & $o$ & $16.43\pm0.02$ \\
155.72 & ATLAS & $o$ & $16.43\pm0.03$ \\
155.73 & ATLAS & $o$ & $16.48\pm0.03$ \\
159.73 & ATLAS & $o$ & $16.60\pm0.03$ \\
159.74 & ATLAS & $o$ & $16.55\pm0.04$ \\
\end{longtable}

\end{appendix}

\end{document}